\PassOptionsToPackage{table}{xcolor}
\documentclass{lightspeed}
\microtypesetup{expansion=false}
\setcitestyle{square,comma,numbers,sort&compress}

\usepackage[T1]{fontenc}
\usepackage[utf8]{inputenc}
\usepackage{latexsym}
\usepackage{microtype}
\usepackage{inconsolata}

\usepackage{booktabs}
\usepackage{multirow}
\usepackage{array}
\usepackage{tabularx}
\usepackage{xcolor}
\usepackage{graphicx}
\usepackage{subcaption}
\usepackage{float}        %
\usepackage{wrapfig}      %
\usepackage{needspace}    %
\usepackage{placeins}     %
\usepackage{capt-of}      %
\usepackage{amsmath}
\usepackage{amssymb}
\usepackage{pgfplots}     %
\pgfplotsset{compat=1.16}
\usepackage{url}

\usepackage{tikz-cd}
\usepackage{multicol}

\renewcommand{\cite}{\citep}

\usepackage[table]{xcolor}   %
\usepackage{bm}              %
\usepackage{enumitem}

\definecolor{mowin}{HTML}{4E8D6E}    %
\definecolor{motie}{HTML}{DCDCDC}    %
\definecolor{moloss}{HTML}{C0655B}   %
\definecolor{bestrow}{HTML}{EAF2EE}  %

\newlength{\prefbarw}
\newlength{\prefsegw}                               %

\newcommand{\prefseg}[4]{%
  \ifnum#2>0
    {\setlength{\fboxsep}{0pt}%
     \setlength{\prefsegw}{\dimexpr #2\prefbarw/#3\relax}%
     \ifdim\prefsegw<0.85em \setlength{\prefsegw}{0.85em}\fi
     \colorbox{#1}{\makebox[\prefsegw][c]{%
       \rule[-0.5ex]{0pt}{1.9ex}%
       \textcolor{#4}{\tiny\bfseries #2}}}%
     \kern 0.5pt}%
  \fi}

\newcommand{\prefbar}[3]{%
  \begingroup
  \edef\ptotal{\number\numexpr#1+#2+#3\relax}%
  \prefseg{mowin}{#1}{\ptotal}{white}%
  \prefseg{motie}{#2}{\ptotal}{black!60}%
  \prefseg{moloss}{#3}{\ptotal}{white}%
  \endgroup}

\newcommand{\legendbox}[1]{%
  {\setlength{\fboxsep}{0pt}\colorbox{#1}{\rule{0pt}{1.6ex}\hspace{1.6ex}}}}

\usepackage{tikz}
\usetikzlibrary{positioning,arrows.meta,fit,backgrounds,calc,shapes.geometric}

\providecommand{\Description}[1]{}

\title{Motion-Omni: End-to-End Joint Speech and Full-Body Motion
  for Spoken Dialogue}

\author[1,*]{Chengqian Ma}
\author[2]{Wei Tao}
\author[3,*]{Haoyu Zhang}
\author[4,\dagger]{Yiwen Guo}

\affiliation[1]{Peking University}
\affiliation[2]{LIGHTSPEED}
\affiliation[3]{The Chinese University of Hong Kong, Shenzhen}
\affiliation[4]{Independent Researcher}
\renewcommand{\affiliationlist}{%
  \affiliationformat[1]{Peking University}\quad
  \affiliationformat[2]{LIGHTSPEED}\par
  \affiliationformat[3]{The Chinese University of Hong Kong, Shenzhen}\quad
  \affiliationformat[4]{Independent Researcher}%
}

\email{machengqian25@stu.pku.edu.cn}
\email{wtao@ieee.org}
\email{haoyuzhang3@link.cuhk.edu.cn}
\email{guoyiwen89@gmail.com}
\renewcommand{\emaillist}{%
  \emailformat{machengqian25@stu.pku.edu.cn}\quad\emailformat{wtao@ieee.org}\\
  \emailformat{haoyuzhang3@link.cuhk.edu.cn}\quad\emailformat{guoyiwen89@gmail.com}%
}

\abstract{An avatar that holds a conversation should decide what to say and to move while saying it, yet these abilities live in separate model families: spoken dialogue models produce speech without motion, and co-speech motion models produce motion only from audio handed to them.
The standard remedy is a cascade that first generates the spoken response and then runs a motion model over the finished audio, which requires a second full inference pass and precludes any joint optimisation between the two.
We present \textbf{Motion-Omni}, an end-to-end framework in which a spoken dialogue model natively outputs explicit facial expression together with hand, upper-body and lower-body motion, generated directly from the hidden states that produce the speech.
Joint training is not optional here: with the speech pathway frozen, motion remains misaligned with the audio, and co-adapting the LLM, Speech Generator and Motion Generator under both objectives is what recovers alignment while retaining spoken-dialogue ability.
Supervision comes from a scalable, model-agnostic pipeline that pseudo-labels consistent-voice speech responses with a replaceable motion teacher, yielding $422{,}856$ quality-ranked pairs ($1{,}402$ hours).
We further release SwDA-500 and, to our knowledge, the first public evaluation protocol for stochastic open-ended full-body spoken dialogue, matching audio across motion systems while unifying rendering, automatic metrics, human evaluation, and latency measurement.
Instantiated with a Qwen2.5-7B-Instruct backbone, \textbf{Motion-Omni-Q7} matches the same-audio teacher cascade to within 2\% on reference-free motion metrics while responding $5.4\times$ faster ($\mathrm{RTF}=0.78$, faster than real time), surpasses all non-teacher cascades on beat correlation and diversity, and reaches a $2.62\%$ word error rate, the lowest among the omni-modal systems compared.
}
\headercontent{%
  \href{https://step-out.github.io/Motion-Omni-Page/}{Project page}\hspace{1.5em}%
  \href{https://github.com/step-out/Motion-Omni}{Code}\hspace{1.5em}%
  \href{https://huggingface.co/datasets/ChengqianMa/Motion-Omni}{Data}%
}
\date{August 2026}

\begin{document}
\thispagestyle{firstheader}
\maketitle
\begingroup
\renewcommand{\thefootnote}{\fnsymbol{footnote}}
\footnotetext[1]{Work is done during internship at LIGHTSPEED.}
\footnotetext[2]{Corresponding author.}
\endgroup
\setcounter{footnote}{0}

\begin{figure}[t!]
\centering
\includegraphics[width=\linewidth]{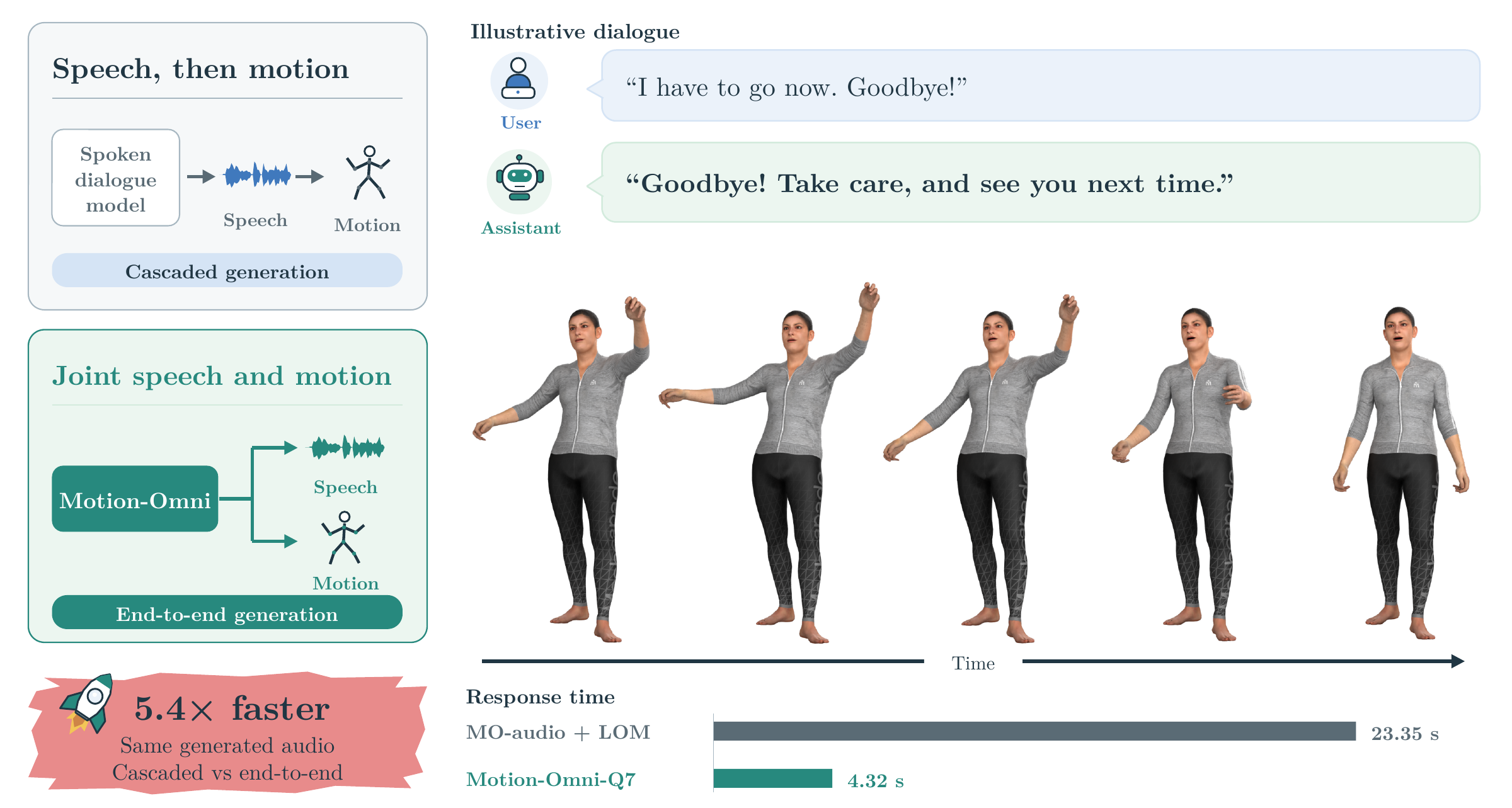}
\caption{Motion-Omni jointly generates speech and full-body motion. The dialogue is illustrative, with unpaired LOM-generated poses. The timing comparison uses the matched-audio, offline SwDA-500 benchmark (Table~\ref{tab:latency}).}
\Description{Left: cascaded and end-to-end generation paradigms. Right: explicitly illustrative farewell dialogue above five LOM-generated poses shown at uniform opacity. A rightward Time arrow indicates chronological order without a quantitative time scale. The poses come from a separate Please-conditioned trial, not Motion-Omni outputs or a measured farewell response. Bottom: separately measured offline response times of 23.35 seconds for MO-audio plus LOM and 4.32 seconds for Motion-Omni-Q7 on SwDA-500.}
\label{fig:teaser}
\end{figure}

\section{Introduction}

Speech and full-body co-speech motion, including facial expression and hand, upper-body, and lower-body movement, are tightly coupled in human communication:
these movements carry meaning that complements and reinforces what is being said.
We call a model that produces both a spoken response and this accompanying motion, conditioned on the same dialogue context, a \emph{spoken motion model}. Figure~\ref{fig:teaser} illustrates this interaction paradigm. Throughout this paper, \emph{motion} denotes co-speech communicative motion; locomotion, dance, sports, and generic action generation are outside our scope.

Spoken dialogue models (SDMs)~\cite{gpt4oaudio,qwen25,llamaomni2024} and co-speech motion generation models~\cite{emage2024,talkshow,lom2024} have each progressed rapidly, and the direct way to obtain both outputs is to run them in sequence: an SDM produces the spoken response, and its audio is then passed to a motion model.
This cascade does produce speech and motion, and it is the baseline we compare against throughout, %
but it has two structural costs: the motion model runs as a second full inference pass after the audio is complete, and no motion objective can ever update the speech or dialogue parameters.
This paper asks whether both costs can be removed at once, that is, whether full-body motion can be a native output of an SDM, generated from the same states that produce the speech, without giving up motion quality or spoken-dialogue ability.
Recent spoken motion models~\cite{solami, umind2026, exomni2025, vibes2025, miburi2026} relax parts of this recipe, but none natively combines facial expression, hand, upper-body, and lower-body output with co-adaptation of the response-generation pathway.

Building and evaluating such a model presents three challenges.
First, making motion a native output of the SDM requires jointly optimising the speech and motion objectives without degrading either output.
Joint optimisation enables reciprocal transfer through shared states: motion supervision can preserve motion-relevant timing and contextual cues in the states that generate speech, while speech supervision constrains those states to retain spoken-dialogue ability~\cite{DBLP:journals/pami/BaltrusaitisAM19}.
Realising this transfer is nontrivial, because the two outputs live at heterogeneous rates ($12.5$\,Hz speech units and $30$\,Hz motion) and share parameters through which the two losses can interfere, so the architecture must bridge the rate mismatch and the training procedure must prevent motion gradients from degrading the dialogue pathway.
Second, such joint training needs supervision at a scale no existing corpus provides.
Many SDMs are trained towards a consistent target voice, so its motion supervision should be paired with responses in that same voice; captured audiovisual corpora record many speakers and voices~\cite{seamlessinteraction} and cannot supply this at scale.
We therefore adopt teacher pseudo-labeling, accepting the teacher as a quality ceiling in exchange for scale and voice consistency, which makes quality ranking of the generated supervision essential.
Third, to our knowledge, no public benchmark is designed for stochastic open-ended full-body spoken dialogue.
Existing full-body motion benchmarks assume fixed supplied speech~\cite{genea2022,miburi2026}, while spoken-dialogue evaluations with controlled generated speech cover facial animation only~\cite{exomni2025}; neither evaluates whether a model's own valid but unpredictable spoken response is accompanied by appropriate whole-body motion.
Fair evaluation therefore requires open-ended prompts, matched audio across motion systems to prevent differences in response content and prosody from confounding motion comparisons, and complementary measurements of speech, motion, human perception, and latency.

In this work we propose \textbf{Motion-Omni}, an end-to-end spoken motion framework.
To our knowledge, it is the first open-ended spoken dialogue model that natively generates explicit facial expression, hand, upper-body, and lower-body motion while allowing the motion objective to update both the LLM and a distinct Speech Generator.
For the architecture challenge, our four-component framework (Speech Projector, LLM Backbone, Speech Generator, Part-Aware Motion Generator) bridges the rate mismatch through a dual-input conditioning interface: the Motion Generator attends to the Speech Generator's hidden states (key/value) and consumes the embeddings of emitted speech tokens (query), interpolated from the speech-unit rate to the motion rate, so motion is generated directly from the representations that produce the speech, without using the rendered waveform as an intermediate input.
A four-stage training progression (Automatic Speech Recognition (ASR), Text-to-Speech (TTS), TTS-with-Motion (TTSM) curriculum, joint mixture) co-adapts the Speech Projector, LLM, Speech Generator and Motion Generator end-to-end while keeping each as a distinct module and preserving spoken-dialogue ability.
An ablation confirms that this co-adaptation is necessary: with the speech pathway frozen, generated motion remains visibly misaligned with the speech (Section~\ref{sec:training}).
For the data challenge, we introduce a scalable, model-agnostic route from open-ended speech-instruction responses in a consistent target voice to explicit facial and full-body motion supervision: a replaceable motion teacher (LOM~\cite{lom2024} in this instantiation) labels every response waveform, and a dual-metric quality score induces the training curriculum. Applied to InstructS2S-200K~\cite{llamaomni2024}, this route yields $422{,}856$ teacher-generated pseudo-labeled speech-motion samples ($1{,}402$ hours).
For the evaluation challenge, we introduce and release SwDA-500 and a reproducible protocol for stochastic open-ended spoken motion: matched generated audio controls response content and prosody, while a shared renderer, reference-aware automatic metrics, paired human evaluation, and a common latency protocol cover the complete output.
Instantiated as \textbf{Motion-Omni-Q7} (denoted as \textbf{MO}) with a Qwen2.5-7B-Instruct~\cite{qwen25} backbone, the resulting model matches the same-audio teacher cascade to within $2\%$ on reference-free motion metrics while completing speech-and-motion responses faster than real time ($\mathrm{RTF}=0.78$), $5.4{\times}$ faster than that cascade; among the remaining systems, which do not run the motion teacher at inference time, it obtains the best beat correlation and diversity, and its speech reaches a $2.62\%$ word error rate, the lowest among the omni-modal systems in our comparison.

\noindent In summary, our contributions are:
(1)~an end-to-end spoken motion framework in which motion is generated from the hidden states that produce the speech, with motion supervision jointly updating the LLM, Speech Generator and Motion Generator; ablations show this co-adaptation is necessary for speech-motion alignment, and the resulting model matches the same-audio teacher cascade on motion quality while removing its separate audio-to-motion stage and responding $5.4{\times}$ faster;
(2)~a scalable, model-agnostic route from consistent-voice speech instructions to quality-ranked facial and full-body motion supervision, yielding $422{,}856$ paired samples ($1{,}402$ hours);
(3)~SwDA-500 together with, to our knowledge, the first publicly released evaluation protocol for stochastic open-ended full-body spoken dialogue, matching audio across motion systems and unifying rendering, automatic metrics, human evaluation, and latency measurement.

\section{Related Work}
\label{sec:related}

Motion-Omni sits at the intersection of four lines of work:
(i)~audio-conditioned co-speech motion generation;
(ii)~text-driven integrated speech-and-gesture synthesis;
(iii)~spoken dialogue models; and
(iv)~dialogue systems that emit both speech and articulated motion.

\subsection{Motion Generation Models}

Co-speech motion generation synthesises body motion from supplied speech audio~\cite{cospeechreview}.
Audio-driven facial animation (FaceFormer~\cite{faceformer}) produces face-only motion with periodic positional encodings that our Motion Generator inherits via Ex-Omni~\cite{exomni2025}.
Full-body co-speech models (TalkShow~\cite{talkshow}, Listen, Denoise, Action~\cite{listendenoiseaction}, EMAGE~\cite{emage2024}, MambaTalk~\cite{mambatalk2024}, GestureLSM~\cite{gesturelsm2025}, SemTalk~\cite{zhang2025semtalk}) extend the recipe to 3D motion using diffusion, discrete, state-space, or flow-matching representations;
the Language of Motion (LOM)~\cite{lom2024} uses four part-specific VQ-VAE codebooks (face, hand, upper, lower) that we reuse as a frozen detokeniser and as the motion teacher in our data pipeline.
These systems can be connected to synthetic speech in a cascade, but they do not themselves plan a dialogue response.

\subsection{Integrated Speech and Gesture Synthesis}

Joint optimisation lets objectives from related modalities shape shared representations, a central form of multimodal co-learning~\cite{DBLP:journals/pami/BaltrusaitisAM19}.
Integrated speech-and-gesture synthesis applies this principle to prescribed text scripts:
\citet{integratedisg2021} adapted neural TTS architectures to jointly predict speech acoustics and 3D gesture from text.
Diff-TTSG~\cite{diffttsg} introduced probabilistic diffusion with parallel speech and gesture heads,
while Match-TTSG~\cite{matchttsg} used one conditional-flow-matching decoder to model their joint distribution.
MAGI~\cite{magi} added synthetic pre-training, multi-speaker support and prosody control.
FastTalker~\cite{fasttalker} reuses intermediate TTS timing and prosodic features for efficient full-body gesture decoding,
while Gelina~\cite{gelina} autoregressively interleaves discrete speech and gesture tokens.
Unlike these systems, which synthesise speech and gesture for a prescribed script, Motion-Omni takes a user turn as input and generates an open-ended spoken response together with its motion.

\subsection{Spoken Dialogue Models}
\label{sec:related_sdm}

Spoken dialogue models (SDMs) equip LLMs with speech input and output.
SpeechGPT~\cite{speechgpt} helped popularise discrete audio tokens as a language-model vocabulary;
LLaMA-Omni~\cite{llamaomni2024} adds a speech-unit decoder and the InstructS2S-200K dataset that we build on;
GLM-4-Voice~\cite{glm4voice} contributes the discrete 12.5\,Hz speech tokenizer and CosyVoice~\cite{cosyvoice}-style flow-matching decoder that MO adopts;
Qwen2.5-Omni~\cite{qwen25omni} introduces a Thinker-Talker architecture;
and Moshi~\cite{moshi} achieves full-duplex dialogue.
More recently, WavAlign~\cite{DBLP:conf/acl/ChenJCLLWWLWPZZ26} proposes modality-aware adaptive post-training for semantic quality and speech expressiveness, while AV-Dialog~\cite{DBLP:conf/acl/ChenVGG26} adds visual cues for target-speaker tracking and turn-taking in noisy multi-speaker settings. These standalone SDMs do not produce body motion.

\subsection{Spoken Motion Models}
\label{sec:related_smm}

The most directly related line conditions joint speech and motion output on an open-ended dialogue context.
\textbf{SOLAMI}~\cite{solami} jointly predicts speech and body/hand motion tokens with an AnyGPT/LLaMA2-based autoregressive backbone, but generates facial animation post hoc with an audio-to-face model.
\textbf{U-Mind}~\cite{umind2026} uses one shared autoregressive backbone to generate response text, acoustic tokens, and SMPL-X pose tokens, but does not specify native facial-expression output or isolate how motion supervision affects speech-generation states.
\textbf{Ex-Omni}~\cite{exomni2025} jointly optimises an LLM, Speech Generator and 52-dimensional ARKit facial decoder, but addresses only the face.
\textbf{ViBES}~\cite{vibes2025} combines a frozen GLM-4-Voice speech expert with trainable face and body experts; freezing preserves its speech model but prevents motion gradients from updating that component.
\textbf{MIBURI}~\cite{miburi2026} causally generates full-body gestures and facial expressions from Moshi's streaming states, but does not report motion-loss adaptation of Moshi. It addresses streaming interaction, whereas Motion-Omni synthesises a complete response offline, so their response times are not directly comparable.
Motion-Omni uniquely combines explicit face, hand, upper- and lower-body output with end-to-end co-adaptation of distinct LLM, speech and motion modules, supported by a model-agnostic pseudo-labeling curriculum.

\begin{figure}[!tp]
\centering
\includegraphics[width=\linewidth]{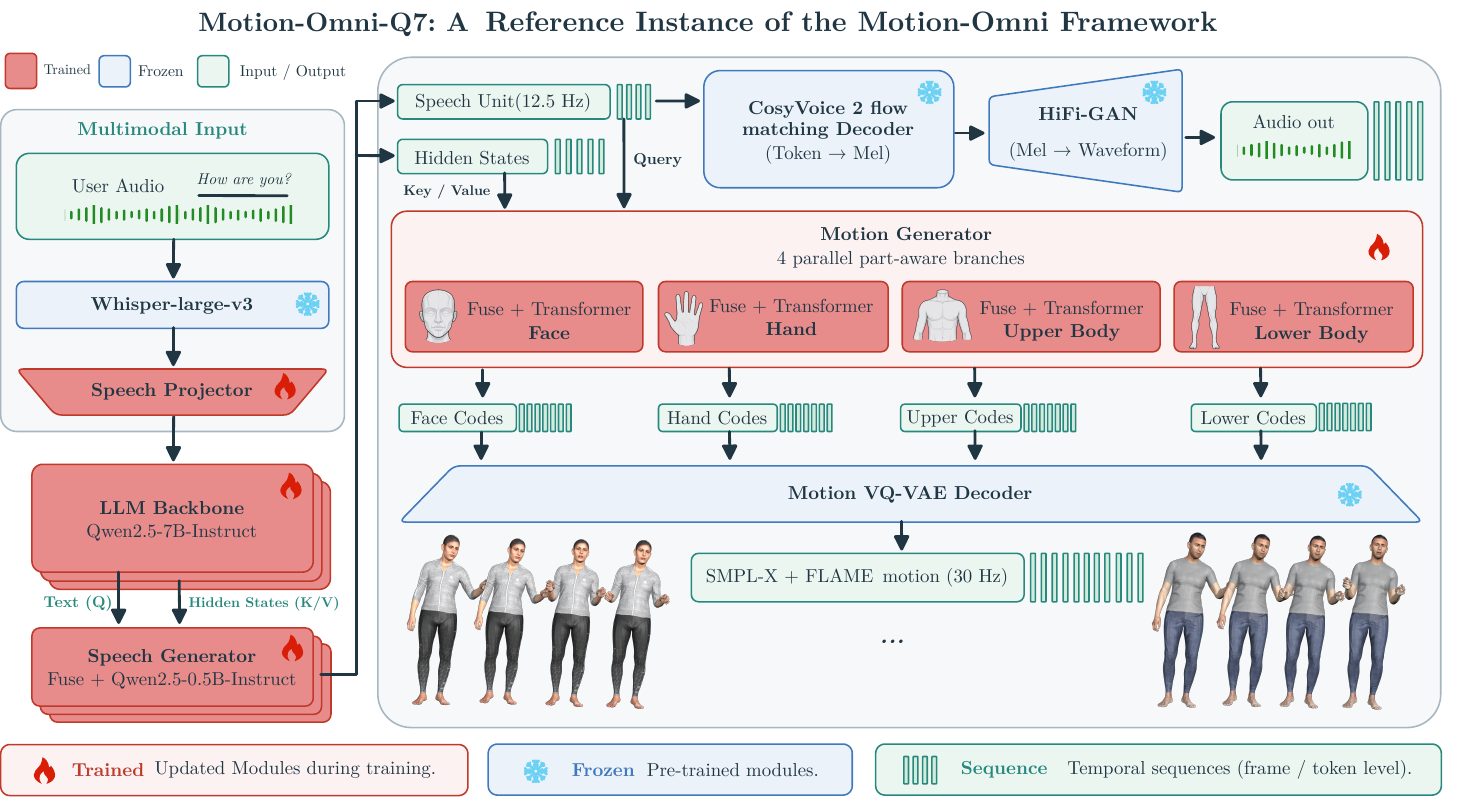}
\caption{Motion-Omni framework with four components and their conditioning topology.}
\label{fig:architecture}
\end{figure}

\section{Method}
\label{sec:method}

\subsection{System Overview}

The Motion-Omni framework specifies four components that together process a user's speech (or text) input and autoregressively generate a spoken response together with synchronised full-body co-speech motion (Figure~\ref{fig:architecture}).
For each component, the framework prescribes only the input/output interface and the conditioning topology; the concrete model class, parameter count and tokenizer are choices made by a particular instance. Motion-Omni-Q7, the reference instance reported here, makes the following choices.
The \textbf{Speech Encoder} is a frozen Whisper-large-v3~\cite{whisper} encoder (hidden dimension 1280) that maps a 16\,kHz waveform to continuous representations; a \emph{speech projector} concatenates every five consecutive frames and passes them through a two-layer MLP into the LLM embedding space, downsampling the sequence fivefold.
The \textbf{LLM Backbone} is a Qwen2.5-7B-Instruct~\cite{qwen25} model. For speech input, projected Whisper features replace a designated \texttt{<speech>} placeholder in the token sequence; the LLM then processes the resulting continuous speech segment and surrounding text tokens. It is frozen during Stages~1--3 and fine-tuned with a small learning rate in Stage~4.
The \textbf{Speech Generator} is a Qwen2-style transformer initialised from Qwen2.5-0.5B-Instruct~\cite{qwen25} that autoregressively emits GLM-4-Voice~\cite{glm4voice} discrete speech units at 12.5\,Hz over a vocabulary of $16{,}384$ units (plus three control tokens). A Token-as-Query Gated Fusion (TQGF) block~\cite{exomni2025} lets token embeddings query the LLM's contextualised hidden states through learned head-wise sigmoid gates; its full formulation is provided in Appendix~\ref{sec:tqgf}.
The \textbf{Motion Generator} comprises four parallel per-part decoders that emit LOM~\cite{lom2024} VQ codes at 30\,Hz for the face, hands, upper body and lower body,
conditioned on the Speech Generator's last-layer hidden states (key/value) and on a learned speech-token query embedding (initialised from the pre-trained flow embedding of CosyVoice~\cite{cosyvoice}).
At inference time, the four part decoders share the Speech Generator context but do not explicitly cross-condition on one another's sampled motion outputs.

At inference time, speech units are converted to mel spectrograms by a CosyVoice~\cite{cosyvoice} chunk-aware flow-matching decoder and then to 22.05\,kHz waveforms by a HiFi-GAN~\cite{hifigan}-style vocoder.
Motion codes are decoded by the frozen LOM VQ-VAE into SMPL-X~\cite{smplx} body/hand parameters and FLAME~\cite{flame2017} facial-expression coefficients. For the human evaluation we render each response as a video using an SMPL-X mesh (Appendix~\ref{sec:supp_render}).

\subsection{Motion Generator}

\paragraph{Motion Generator.}
The Motion Generator instantiates one independent decoder per body part $b\in\{\mathrm{face},\allowbreak\mathrm{hand},\allowbreak\mathrm{upper},\allowbreak\mathrm{lower}\}$. Each decoder stacks $L_\text{TQGF}=2$ TQGF layers, which reuse Equations~\eqref{eq:tqgf_gate}--\eqref{eq:tqgf_res} at $d_m=512$, $h=8$ and $d_h=64$, followed by an $L_\text{self}=6$-layer self-attention Transformer with periodic rotary positional encoding~\cite{exomni2025} of period $\mathcal{T}=30$, that is, one second at $30$\,Hz. Its two input streams are
\begin{align}
  \mathbf{Z} &= \mathbf{H}_s\mathbf{W}_h,
    \label{eq:motion_kv}\\
  \mathbf{Q} &= \mathrm{Interp}\big(\mathbf{E}[\mathbf{u}]\mathbf{W}_e\big),
    \label{eq:motion_q}
\end{align}
where $\mathbf{H}_s$ are the Speech Generator's last-layer hidden states and $\mathbf{W}_h\in\mathbb{R}^{896\times d_m}$ projects them into the motion working dimension. The sequence $\mathbf{u}\in\{0,\dots,16{,}383\}^{T_s}$ holds speech units, $\mathbf{E}$ is a $512$-dimensional look-up table initialised from the pre-trained CosyVoice flow embedding~\cite{cosyvoice}, $\mathbf{W}_e$ is a learned projection, and $\mathrm{Interp}$ linearly interpolates along the time axis from the $12.5$\,Hz speech-unit rate to the $30$\,Hz motion rate. All four decoders read the same $\mathbf{Z}$ and $\mathbf{Q}$ and differ only in their parameters. The self-attention block that follows operates at motion rate under a causal mask, so each frame attends only to earlier frames. During training $\mathbf{u}$ are the ground-truth units of the target response; at inference they are the units the Speech Generator has just emitted.

Conditioning on $\mathbf{H}_s$ in Equation~\eqref{eq:motion_kv} rather than on the decoded waveform is what removes a separate audio-to-motion stage at inference time, and it gives each decoder a representation that already carries acoustic timing together with response semantics. The same query/key asymmetry holds here: Equation~\eqref{eq:motion_q} supplies one embedding per speech unit, identifying what is being said at that instant, while the keys and values additionally carry speaker timbre, which does not drive body movement, and the gate selects what each body part needs. A part-specific MLP head then produces per-frame logits over the $|\mathcal{C}|=256$ LOM codebook entries, trained with
\begin{equation}
  \mathcal{L}_\text{motion} = \sum_b w_b\,\mathcal{L}^{(b)}_\text{CE},
  \qquad
  w_\text{face}\!:\!w_\text{hand}\!:\!w_\text{upper}\!:\!w_\text{lower}=106\!:\!180\!:\!78\!:\!61,
  \label{eq:motion_loss}
\end{equation}
where $\mathcal{L}^{(b)}_\text{CE}$ is the per-frame cross-entropy for part $b$ with label smoothing $0.1$, and the weights $w_b$ are normalised to sum to one and are proportional to the underlying SMPL-X+FLAME feature dimensions, so that every VQ-code prediction carries equal per-feature-dimension importance. Full architecture and implementation details are listed in Appendix~\ref{sec:arch_details}.

\subsection{Progressive Training Curriculum}
\label{sec:training}

We train Motion-Omni-Q7 in four stages; throughout, the Whisper speech encoder and the LOM VQ-VAE are frozen.
\textbf{Stage~1} trains only the speech projector with ASR supervision. Each sample contains a \texttt{<speech>} placeholder followed by its transcript target; projected Whisper features replace the placeholder, labels at the inserted speech positions are masked, and next-token cross-entropy is applied only to the transcript tokens. Because the LLM is frozen, this objective trains the projector to produce representations from which the LLM can decode the transcript.
\textbf{Stage~2} trains the Speech Generator on TTS-style pairs while the LLM remains frozen. For example, the text-side instruction asks the model to say or restate a supplied sentence, the LLM produces a semantically consistent response representation, and the target is the corresponding sequence of discrete speech units.
\textbf{Stage~3} attaches the Motion Generator and jointly trains it with the Speech Generator, while keeping the LLM and projector frozen, on the TTSM corpus produced by our data construction pipeline (Section~\ref{sec:dataset}); the data is consumed under a four-substage curriculum that exposes the network to progressively larger quality quantiles in turn, warm-starting each substage from the previous one. In an initial pilot that kept the Speech Generator frozen, the motion loss plateaued above the level reached by joint training, and the rendered motions were clearly misaligned with the speech audio. Jointly updating the Speech Generator lowered the loss further, improved audio/motion alignment, and is therefore used throughout Stages~3a--3d.
\textbf{Stage~4} unfreezes the LLM backbone, the speech projector, the Speech Generator and the Motion Generator simultaneously and optimises a four-task ASR/TTS/Speech-to-Speech-with-Motion (S2SM)/T2T mixture with a constant-with-warmup schedule, warm-started from the final Stage~3d checkpoint.
Numerical-stability measures and per-stage hyperparameters are detailed in Appendix~\ref{sec:supp_training_hparams}.

\subsection{Data Construction Pipeline}
\label{sec:dataset}

\begin{figure}[H]
\centering
\includegraphics[width=\linewidth]{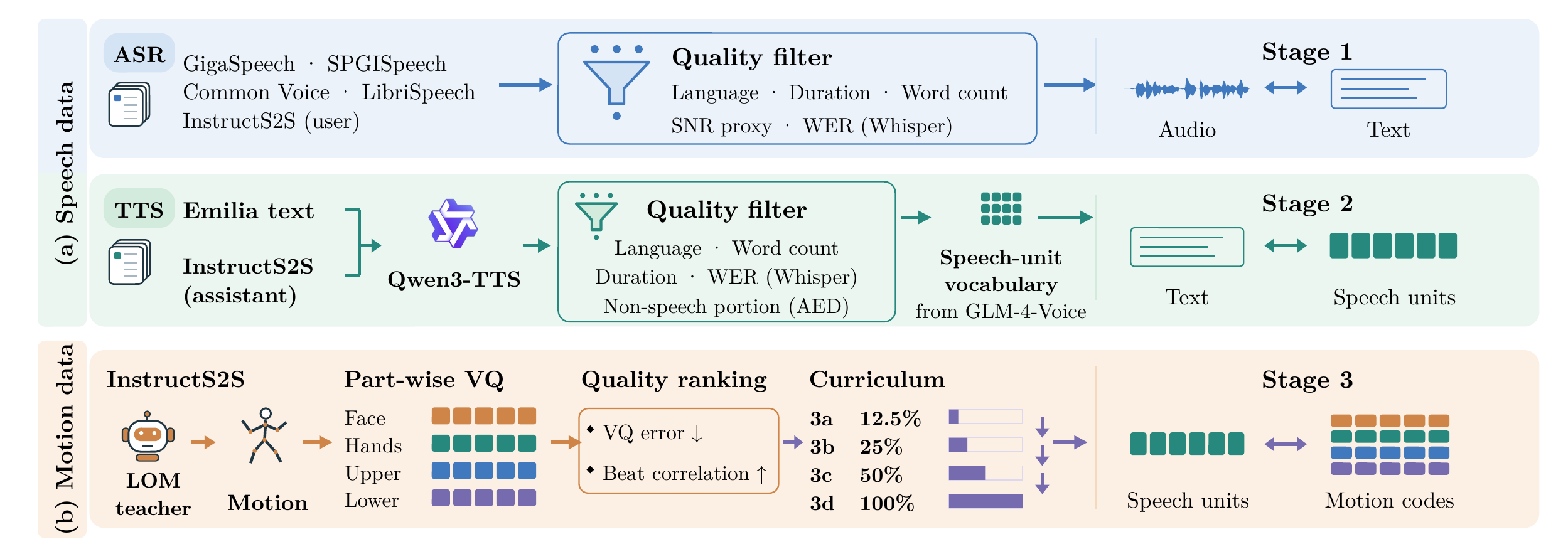}
\caption{Speech-data curation and motion supervision. Insets schematically show the paired modalities: audio--text in Stage~1, text--speech units in Stage~2, and speech units--motion codes in Stage~3.}
\Description{Two speech-data workflows and a motion-supervision workflow with directional process arrows and double-headed pairing arrows. Both text sources pass through Qwen3-TTS, marked by its official logo; a vector-block icon represents the GLM-4-Voice vocabulary. All token cells have uniform solid fill within each stream. Stage 3 retains a continuous 12.5, 25, 50 and 100 percent curriculum.}
\label{fig:data_pipeline}
\end{figure}

Figure~\ref{fig:data_pipeline} summarises speech-data preparation and the construction of motion supervision.

Training a spoken motion model requires large-scale supervision that pairs a dialogue response with the motion accompanying it. Captured audiovisual resources such as the dyadic Seamless Interaction corpus~\cite{seamlessinteraction} record conversations between speakers, but their many voices are not matched to the single target voice our Speech Generator is trained to produce. We therefore construct supervision by pseudo-labeling: a \emph{generic} pipeline takes a speech-instruction corpus already paired with response audio together with a co-speech motion generation teacher, and produces paired (text, speech, full-body motion) supervision in two steps.
\begingroup
\setlength{\columnsep}{6pt}
\setlength{\intextsep}{4pt}
\begin{wraptable}{r}{0.30\textwidth}
\centering\scriptsize
\setlength{\abovecaptionskip}{0pt}
\caption{Training data by stage.}
\label{tab:training}
\setlength{\tabcolsep}{5pt}
\begin{tabular}{@{}lrr@{}}
\toprule
\textbf{Stage} & \textbf{Samples} & \textbf{Hours} \\
\midrule
Stage~1: ASR     & 1,572,119 & 4,295.8 \\
Stage~2: TTS     & 1,642,715 & 4,517.0 \\
Stage~3: TTSM    & 422,856   & 1,402.3 \\
Stage~4: Mixture & 724,508   & 1,929.4 \\
\midrule
Total              & 4,362,198 & 12,144.6 \\
\bottomrule
\end{tabular}
\end{wraptable}
\textbf{(1) Motion supervision via a teacher}: a pre-trained co-speech motion generation teacher (LOM~\cite{lom2024} in our run) is run on every response waveform; its four per-part VQ code streams form the motion target. The teacher is a replaceable component, so a future implementation can regenerate supervision with a stronger teacher. \textbf{(2) Dual-metric quality scoring and curriculum}:
every sample is scored along
(i)~a \emph{weighted VQ-VAE reconstruction error} $L_1^\text{vq}$ that flags motion outside the codebook's expressive range,
and (ii)~a \emph{beat correlation} (BC) score~\cite{DBLP:conf/iccv/LiYRK21} that flags teacher motion weakly coupled to the speech;
after robust 5th--95th percentile normalisation each sample receives a combined score $s(x)= \alpha(1-\widetilde{L_1^\text{vq}}) + (1-\alpha)\widetilde{\mathrm{BC}}$ with $\alpha=0.5$, which drives the four-substage Stage~3 curriculum and selects the Stage~4 S2SM pool. The exact reconstruction-error computation is given in Appendix~\ref{sec:vqvae_error}.
\par\endgroup

\paragraph{Reference instantiation.}
Table~\ref{tab:training} summarises the per-stage data statistics.
\textbf{Stages~1--2} draw ASR and TTS pairs from InstructS2S-200K~\cite{llamaomni2024} and the English subset of Ex-Instruct~\cite{exomni2025}.
\textbf{Stage~3} applies the pipeline to InstructS2S-200K with the LOM teacher. Training directly on the complete unranked corpus diverged, so we instead expose the network to progressively larger quantiles of the dual-metric score ($12.5\%/25\%/50\%/100\%$), warm-starting each substage from the previous checkpoint. The stage-wise teacher-reference Fr\'{e}chet Gesture Distance (FGD, defined in Section~\ref{sec:motion_results}) reported in Appendix~\ref{sec:supp_ablations} decreases from $0.3974$ at Stage~3a to $0.3040$.
\textbf{Stage~4} assembles a four-task ASR/TTS/S2SM/T2T mixture. The T2T component (${\sim}213$K samples) is drawn from six text-only datasets (SODA~\cite{soda}, WildChat~\cite{wildchat}, Tulu-3~\cite{tulu3}, WizardLM~\cite{wizardlm}, OpenThoughts~\cite{openthoughts}, NuminaMath-CoT~\cite{numinamath}) to preserve general text-response planning and multi-turn dialogue behaviour during joint fine-tuning. Without this component, responses more often repeated the user's input.
All sources are English-filtered with \texttt{langdetect} and capped at $10{,}000$ characters per dialogue. Further pipeline details are provided in Appendix~\ref{sec:data_details}.

\section{Experiments}

\subsection{Experimental Setup}
\label{sec:experimental_setup}

For speech and motion evaluation we use \emph{SwDA-500}, an external 500-prompt dialogue-text evaluation set derived from the Switchboard Dialog Act Corpus (SwDA)~\cite{swda}. The set covers all 66 SwDA topic descriptions with 7 or 8 semantically complete speaker turns per topic, keeps prompts of moderate length, and removes transcription artifacts not intended to be spoken. SwDA-500 provides real conversational wording, but it is not a paired real-motion benchmark; all motion references below are teacher-generated or baseline-generated under matched prompts.

\Needspace{21\baselineskip}
\subsection{Spoken Dialogue Quality}
\label{sec:voicebench}

\begingroup
\setlength{\columnsep}{6pt}
\setlength{\intextsep}{0pt}
\begin{wraptable}[12]{r}{0.30\textwidth}
\centering\scriptsize
\setlength{\tabcolsep}{4pt}
\captionsetup{font=scriptsize,skip=4pt}
\caption{WER (\%) on Seed-TTS-Eval.}
\label{tab:speech}
\begin{tabular}{lc}
\toprule
\textbf{Model} & \textbf{test-en}$\downarrow$ \\
\midrule
\multicolumn{2}{c}{\cellcolor[rgb]{0.88,0.95,1}\textbf{Human Reference}} \\
\addlinespace[1pt]
Human                                            & 2.14 \\
\addlinespace[1pt]
\multicolumn{2}{c}{\cellcolor[rgb]{0.88,0.95,1}\textbf{Dedicated TTS}} \\
\addlinespace[1pt]
CosyVoice~\cite{DBLP:journals/corr/abs-2407-05407}                       & 4.29 \\
CosyVoice 2~\cite{cosyvoice}                     & 2.57 \\
FireRedTTS~\cite{fireredtts}                      & 3.82 \\
Qwen3-TTS-12Hz-1.7B~\cite{qwen3tts}               & \textbf{1.24} \\
Qwen3-TTS-12Hz-0.6B~\cite{qwen3tts}               & \underline{1.32} \\
CosyVoice 3-1.5B RL~\cite{cosyvoice3}             & 1.45 \\
F5-TTS~\cite{f5tts}                               & 2.04 \\
IndexTTS2~\cite{indextts2}                        & 2.18 \\
\addlinespace[1pt]
\multicolumn{2}{c}{\cellcolor[rgb]{0.88,0.95,1}\textbf{Omni-modal LLMs}} \\
\addlinespace[1pt]
Qwen2.5-Omni~\cite{qwen25omni}                   & 2.72 \\
Ex-Omni~\cite{exomni2025}                        & \underline{2.67} \\
Motion-Omni-Q7 (ours)                            & \textbf{2.62} \\
\bottomrule
\end{tabular}
\end{wraptable}

The remaining evaluation assumes that the backbone can still hold a spoken conversation after joint motion training. We verify this on VoiceBench~\cite{voicebench}, a nine-subset benchmark covering open-ended dialogue, factual QA, reasoning, instruction-following and safety, where Motion-Omni-Q7 reaches an Overall of $47.63$, above LLaMA-Omni ($41.12$), Ex-Omni ($43.57$), Mini-Omni2 ($33.49$) and Moshi ($29.51$). Appendix~\ref{sec:supp_voicebench} reports the per-subset scores for all systems.

\subsection{Text-to-Speech (Seed-TTS-Eval)}
\label{sec:speech_results}

\paragraph{Setting.}
We evaluate speech intelligibility on Seed-TTS-Eval~\cite{seedtts} (English split, 1{,}088 samples), reporting corpus-level Word Error Rate (WER) computed by Whisper-large-v3~\cite{whisper} on the model's synthesised speech against the ground-truth transcript.

\paragraph{Results.}
Table~\ref{tab:speech} reports WER. In the table, lower is better. \textbf{Bold} marks the best value and \underline{underline} the second best \emph{within each system category}, since dedicated TTS systems synthesise a supplied sentence whereas omni-modal LLMs also plan the response; the Human row is a reference and is not ranked. Motion-Omni-Q7 reaches $2.62\%$ on \texttt{test-en}, the lowest among the omni-modal LLMs, while jointly producing speech and SMPL-X/FLAME co-speech motion; several recent dedicated TTS systems report lower WER on this speech-only metric.

\paragraph{Speech naturalness proxy.}
Because WER only measures intelligibility, we also report UTMOSv2~\cite{utmosv2} as an automatic proxy for speech naturalness on SwDA-500 generated audio, where Motion-Omni-Q7 scores $3.77$: above GLM-TTS, VoxCPM1.5, F5-TTS and CosyVoice, and below Qwen3-TTS and CosyVoice~3. UTMOSv2 does not replace a listening test, so we treat it as a lightweight check under conversational wording and defer the full table to Appendix~\ref{sec:supp_utmos}.

\par\endgroup

\subsection{Speech-to-Motion}
\label{sec:motion_results}

We compare MO against controlled cascades that combine either MO audio or Qwen2.5-Omni~\cite{qwen25omni} audio with recent or previously used audio-to-motion generators: MambaTalk~\cite{mambatalk2024}, GestureLSM~\cite{gesturelsm2025}, EMAGE~\cite{emage2024}, and LOM~\cite{lom2024}. Throughout, a cascade is named \emph{speech source} + \emph{audio-to-motion model}, in the order the two run. Rows using MO audio isolate the motion-generation pathway under the same speech as our model; rows using Qwen2.5-Omni audio test the same motion generators under a stronger external speech model. LOM rows are teacher-reference cascades because LOM also supplies the reference distribution for teacher-based FGD.

\subsubsection{Automatic Evaluation}

\paragraph{Setting.}
Following LOM~\cite{lom2024}, we report FGD~\cite{DBLP:journals/tog/YoonCLJLKL20}, which compares the distribution of generated motion features against a reference distribution, together with Diversity~\cite{DBLP:conf/iccv/0071KPZZ0B21} and BC~\cite{DBLP:conf/iccv/LiYRK21}. Because MO generates speech and motion jointly, we cannot fix its audio output to match a standard benchmark corpus (e.g.\ BEAT2); instead we use LOM's predictions on the \emph{same} generated audio as the FGD reference. This is a teacher-reference FGD: it measures fidelity to the LOM-generated teacher distribution. We omit FGD for LOM rows because their motion supplies the reference distribution. The exploratory $\mathrm{BC}_\text{bi}$ is defined and analysed in Appendix~\ref{sec:bc_pipe}.

\paragraph{Facial geometry and lip synchronisation.}
For facial geometry, we compute the mean squared error (MSE) and the landmark velocity difference (LVD) on facial vertices generated from jaw pose and FLAME expression coefficients, using the LOM-generated motion as the pseudo-reference; these two metrics therefore measure fidelity to the LOM teacher facial-motion distribution rather than direct agreement with newly captured real facial motion. For lip synchronisation, we render the generated SMPL-X+FLAME motion into standardised frontal face videos and evaluate them with the lip-sync error scripts released in the official Wav2Lip evaluation code~\cite{wav2lip}, which score a rendered video against its audio with a pretrained SyncNet and report a confidence (LSE-C, higher is better) and a feature distance (LSE-D, lower is better). Neither needs a motion reference. All systems use the same renderer, avatar, face crop, 25 FPS video, and 16\,kHz mono audio.

\paragraph{Results.}
Table~\ref{tab:motion} compares all systems on SwDA-500 under matched prompts and a single shared metric pipeline. Motion-Omni-Q7 places first or second on seven of the eight metrics. Among systems that do not run LOM at motion-inference time it obtains the highest beat correlation ($7.59$), the highest diversity ($13.67$) and the best score on every facial and lip-sync metric; the rows that surpass it are the LOM teacher-reference cascades, which invoke at motion-inference time the same model that supplied its training targets. It also obtains the lowest teacher-reference FGD, which by construction measures distance to the LOM-generated distribution.

\begin{table}[!tb]
\centering\scriptsize
\caption{Motion, facial geometry and lip-sync metrics on SwDA-500. Teacher-referenced metrics compare against LOM-generated motion and thus favour LOM-based cascades; FGD, BC and $\mathrm{BC}_\text{bi}$ are in ${\times}10^{-1}$ units. \textbf{Bold}/\underline{underline} mark the best/second-best; dashes denote values not computed.}
\label{tab:motion}
\begin{tabular*}{\textwidth}{@{\extracolsep{\fill}}lcccccccc@{}}
\toprule
& \multicolumn{5}{c}{\textbf{Reference-free}} & \multicolumn{3}{c}{\textbf{Teacher-referenced}} \\
\cmidrule(lr){2-6}\cmidrule(lr){7-9}
\textbf{System} & \textbf{Div.}$\uparrow$ & \textbf{BC}$\uparrow$ & $\mathbf{BC_\text{bi}}\uparrow$ & \textbf{LSE-C}$\uparrow$ & \textbf{LSE-D}$\downarrow$ & \textbf{FGD}$\downarrow$ & \textbf{MSE}$\downarrow$ & \textbf{LVD}$\downarrow$ \\
\midrule
Qwen2.5-Omni + MambaTalk  & 11.92 & 7.17 & 8.56 & 6.853 & 7.920 & 4.08 & 7.092 & 6.851 \\
MO-audio + MambaTalk      & 12.07 & 7.03 & 8.60 & 6.797 & 7.955 & 4.15 & 7.128 & 7.019 \\
Qwen2.5-Omni + GestureLSM & 13.61 & 7.44 & 9.04 & -- & -- & 3.29 & -- & -- \\
MO-audio + GestureLSM     & 13.55 & 7.49 & \underline{9.11} & -- & -- & \underline{3.17} & -- & -- \\
Qwen2.5-Omni + EMAGE      & 11.67 & 7.31 & 8.66 & 6.578 & 8.151 & 3.75 & 7.553 & 7.205 \\
MO-audio + EMAGE          & 10.95 & 7.32 & 8.73 & 6.643 & 8.094 & 3.29 & 7.627 & 7.291 \\
\midrule
Qwen2.5-Omni + LOM        & \textbf{13.82} & 7.56 & \textbf{9.13} & \textbf{7.136} & 7.128 & -- & \textbf{6.581} & \textbf{6.243} \\
MO-audio + LOM            & 13.07 & \textbf{7.67} & 9.05 & 6.985 & \textbf{7.011} & -- & -- & -- \\
\midrule
Motion-Omni-Q7 (ours)   & \underline{13.67} & \underline{7.59} & 9.07 & \underline{7.011} & \underline{7.023} & \textbf{3.03} & \underline{6.723} & \underline{6.427} \\
\bottomrule
\end{tabular*}
\end{table}

The integrated model matches strong cascades in motion quality while avoiding a separate audio-to-motion inference stage, and Section~\ref{sec:latency} quantifies what that stage costs.

\begin{table}[!tb]
\centering\scriptsize
\caption{Human preference on motion-related dimensions. \textbf{Bold}/\underline{underline} mark the largest/second-largest pooled wins-minus-losses margin.}
\label{tab:human_pref}
\begin{tabular*}{\textwidth}{@{\extracolsep{\fill}}lcccc@{}}
\toprule
\textbf{Comparison} & \textbf{R1 Rhythm} & \textbf{R2 Semantics} & \textbf{R3 Naturalness} & \textbf{Pooled W/T/L (W--L)} \\
\midrule
Qwen2.5-Omni + LOM
& \prefbar{12}{3}{10} & \prefbar{13}{0}{12} & \prefbar{10}{3}{12} & 35/6/34 ($+1$) \\
Qwen2.5-Omni + EMAGE
& \prefbar{7}{2}{16} & \prefbar{10}{3}{12} & \prefbar{16}{4}{5} & 33/9/33 ($0$) \\
\midrule
\multicolumn{5}{@{}l}{\itshape Same MO speech audio:} \\[1pt]
MO-audio + LOM
& \prefbar{5}{12}{8} & \prefbar{9}{8}{8} & \prefbar{11}{7}{7} & 25/27/23 (\underline{$+2$}) \\
MO-audio + EMAGE
& \prefbar{13}{7}{5} & \prefbar{15}{3}{7} & \prefbar{17}{0}{8} & \textbf{45/10/20} ($\bm{+25}$) \\
\bottomrule
\end{tabular*}
\vspace{2pt}
\begin{flushleft}
\footnotesize
\legendbox{mowin}~MO wins \quad
\legendbox{motie}~ties \quad
\legendbox{moloss}~MO losses
\end{flushleft}
\end{table}

\FloatBarrier
\subsubsection{Human Evaluation}

\paragraph{Setting.}
Prior video generation work has used 3 expert annotators for evaluation~\cite{Long_2026_CVPR}. Here, four trained young-adult male annotators (mean age approximately 26), all fluent in English and experienced in evaluating motion-generation or spoken-dialogue outputs, independently rated 25 paired clips for each of the four arms in Table~\ref{tab:human_pref}. Before annotation, all annotators received the rubric in Appendix~\ref{sec:supp_rubric}. The interface hid system names and randomised A/B order; Appendix~\ref{sec:supp_human_pref} gives the full protocol.
Each pair was rated on three motion-related dimensions: \textbf{R1}~rhythm, \textbf{R2}~semantic alignment, and \textbf{R3}~body naturalness, judged from rendered video+audio. Per-pair verdicts were determined by majority vote.

\paragraph{Results.}
Table~\ref{tab:human_pref} reports win/tie/loss for Motion-Omni-Q7. The two MO-audio arms hold speech fixed, so differences there reflect the motion pathway alone. Against MO-audio + EMAGE the pooled margin is $+25$ ($45$ wins, $10$ ties, $20$ losses). Against MO-audio + LOM, that is, against the teacher itself, the outcome is $25/27/23$, with ties the single most frequent verdict. With four annotators and 25 pairs per arm, we do not read the smaller margins as evidence of a reliable preference.

The same-audio comparisons suggest that Motion-Omni-Q7 can produce motion preferred over EMAGE and broadly comparable to LOM in this limited setting; representative outputs and matched-audio visualisations are provided in Appendix~\ref{sec:qualitative}.
We attribute this to a domain mismatch in the baselines' training data. The training corpus of LOM and EMAGE (BEAT2~\cite{emage2024}) contains not only conversational clips but also monologue and public-speaking recordings; as a result, their generated motion sometimes exhibits presentation-style behaviours (pacing, turning sideways, exaggerated arm swings) that annotators perceive as unnatural in a dialogue setting.
MO, by contrast, is trained on dialogue-paired pseudo-motion produced by our data construction pipeline (Section~\ref{sec:dataset}), so its motion style is calibrated to a conversational register. This may contribute to the observed R3 preference over EMAGE under identical speech.

\FloatBarrier
\subsection{Latency}
\label{sec:latency}

\paragraph{Setting.}
We measure total response time $T_\text{resp}$ (wall-clock seconds from receiving the user audio to completing all audio and motion generation) and the real-time factor $\mathrm{RTF}=T_\text{resp}/T_\text{user}$,
where $T_\text{user}$ is the user-utterance duration; $\mathrm{RTF}<1$ means the system responds faster than real time.
All measurements use the same SwDA-500 evaluation set on a single GPU. These measurements evaluate offline response generation.

\begin{table}[!tb]
\centering\scriptsize
\caption{Response latency on SwDA-500. \textbf{Bold}/\underline{underline} mark the best/second-best value; Slowdown is relative to our model.}
\label{tab:latency}
\begin{tabular*}{\textwidth}{@{\extracolsep{\fill}}lccc@{}}
\toprule
\textbf{Model} & $T_\text{resp}$(s)$\downarrow$ & RTF$\downarrow$ & Slowdown \\
\midrule
Qwen2.5-Omni + LOM   & 99.08 & 18.34 & $22.9{\times}$ \\
Qwen2.5-Omni + EMAGE & 48.65 & 8.79  & $11.3{\times}$ \\
\midrule
MO-audio + LOM       & 23.35 & 4.39  & $5.4{\times}$ \\
MO-audio + EMAGE     & \underline{4.63}  & \underline{0.84}  & $1.1{\times}$ \\
\midrule
Motion-Omni-Q7 (ours)    & \textbf{4.32} & \textbf{0.78} & $1.0{\times}$ \\
\bottomrule
\end{tabular*}
\end{table}

\paragraph{Results.}
Table~\ref{tab:latency} shows that Motion-Omni-Q7 completes a full response (speech + motion) in $4.32\,\mathrm{s}$ ($\mathrm{RTF}=0.78$), faster than real time.
The matched comparison is MO-audio + LOM, which uses the same speech model and the teacher motion model but invokes the latter as a separate audio-to-motion stage: at $23.35\,\mathrm{s}$ it is $5.4{\times}$ slower.

Figure~\ref{fig:quality_latency} in Appendix~\ref{sec:supp_quality_latency} places the same systems on the two axes together. Two cascades come close to our model on one axis each, and neither does so on both: MO-audio + EMAGE comes close in response time but gives up $0.27$ BC, while MO-audio + LOM exceeds its BC by $0.08$ at $5.4{\times}$ the response time.

\FloatBarrier

\section{Conclusion}

We have presented Motion-Omni, an end-to-end framework that enables a conversational LLM to jointly generate intelligible speech and synchronised full-body co-speech motion, together with a model-agnostic pseudo-labeling pipeline that supplies $1{,}402$ hours of teacher-generated supervision and a reproducible evaluation setup for this task class. Across the reported benchmarks, the reference instance Motion-Omni-Q7 records the lowest word error rate among the omni-modal systems compared, and on SwDA-500 the highest beat correlation and diversity among systems that do not run the motion teacher at inference time, along with better rendered-video lip synchronisation than the evaluated EMAGE and MambaTalk cascades.
Conditioning motion on the Speech Generator's hidden states rather than on a decoded waveform keeps the reference-free motion metrics within about $1\%$ of the same-audio teacher cascade while removing a separate audio-to-motion stage, which accounts for the $5.4{\times}$ difference in response time.
All components are swappable; additional instances follow from re-running the recipe with stronger backbones or teachers.

\section*{Limitations}
\label{sec:limitations}
Motion quality is bounded by the LOM~\cite{lom2024} VQ-VAE codebook and teacher-generated pseudo-labels used in this instantiation, so motion outside that distribution cannot be expressed;
we have trained only the Motion-Omni-Q7 instance and leave stronger backbones, larger co-speech motion generation teachers and continuous motion heads to future work.
The current model does not condition on explicit speaker identity or emotion, and our training data is entirely in English, which restricts generalisation to other languages and motion cultures.
Motion-Omni-Q7 ingests the user utterance fully before emitting the first response token, so the system is an offline response generator rather than a streaming interaction model; achieving low-latency interactive behaviour as in streaming systems such as MIBURI~\cite{miburi2026} requires a different design and evaluation protocol.
Existing automatic motion metrics (FGD, BC) are imperfect proxies for perceived naturalness; the current four-annotator A/B/tie human preference comparison is exploratory, and the absence of a larger non-author human study remains a limitation.
We initially attempted to scale rubric scoring with a video-input LLM-as-judge on a broader pilot rubric, but the judge correlates with human raters only on speech quality, so the paper relies on human ratings for perceptual observations (Appendix~\ref{sec:supp_videojudge}).

\subsection*{Reproducibility statement}

The architecture and its conditioning topology are specified in Section~\ref{sec:method}, and the four training stages are described in Section~\ref{sec:training}, with per-stage hyperparameters, optimiser settings and numerical-stability measures given in Appendix~\ref{sec:supp_training_hparams}. The data construction pipeline, including the teacher model, the two quality metrics and the curriculum quantiles, is described in Section~\ref{sec:dataset}, with the exact definitions of the weighted VQ-VAE reconstruction error and the beat correlation score in Appendix~\ref{sec:vqvae_error} and Appendix~\ref{sec:bc_pipe}; all source corpora are public. The evaluation setup is described in Section~\ref{sec:experimental_setup}: SwDA-500 is derived from the public Switchboard Dialog Act Corpus by the filtering procedure stated there, cascade baselines are driven by audio generated by our own model so that speech is held fixed across systems, and every system is rendered through the shared pipeline detailed in Appendix~\ref{sec:supp_render} before lip-sync scoring. The human evaluation protocol, annotator instructions and rubric are given in Appendix~\ref{sec:supp_human_pref} and Appendix~\ref{sec:supp_rubric}. %

\subsection*{Ethics statement}

This work involves a small human preference study. Annotators rated pairs of rendered animation clips with anonymised system labels and randomised presentation order; the protocol is described in Appendix~\ref{sec:supp_human_pref}. No personal data were collected from annotators, and the evaluation prompts are drawn from the public Switchboard Dialog Act Corpus rather than from newly recorded human subjects. The model generates a synthetic speaking avatar with a single fixed voice and a generic body mesh; it is not designed to imitate a specific person's voice, face or motion style, and it does not condition on speaker identity. Training data are entirely in English, so the learned motion style reflects the conversational register of that data and should not be assumed to transfer across languages or cultures.

\subsection*{AI use statement}

We used generative AI tools as writing assistants: to polish phrasing, tighten wording and check the internal consistency of tables and cross-references in the manuscript. Research ideation, the system design, the implementation, the experiments and the analysis of results were carried out by the authors. AI tools were also used for routine coding assistance during development; all such code was reviewed and tested by the authors. AI assistance also supported schematic layouts and figure-generation code; the displayed avatar frames and audio waveforms retain their model-output provenance, including the LOM-generated poses in the illustrative teaser. AI assistants were not used to generate or alter experimental results or metric values, and were not used in the human evaluation. We have reviewed all AI-assisted content and take full responsibility for the final content of this work, including its text, claims and artifacts.

\bibliography{references}
\bibliographystyle{plainnat}

\appendix
\section{Model Architecture Details}
\label{sec:arch_details}

\subsection{Speech Generator}
\label{sec:tqgf}
The Speech Generator is a Qwen2-style transformer initialised from Qwen2.5-0.5B-Instruct~\cite{qwen25} and trained under a next-token-prediction objective over the GLM-4-Voice~\cite{glm4voice} discrete speech-unit vocabulary. It runs at width $d=896$ while the LLM backbone emits $3584$-dimensional states, so the two are joined by a two-layer GELU-MLP and a reshape that together form the conditioning inputs of a Token-as-Query Gated Fusion (TQGF) block~\cite{exomni2025}:
\begin{equation}
  \mathbf{Z}=\mathrm{Reshape}\big(\mathrm{MLP}(\mathbf{H}_\text{llm})\big)\in\mathbb{R}^{4T_t\times d},
  \qquad
  \mathbf{Q}=\mathbf{E}_\text{sg}[\mathbf{w}]\in\mathbb{R}^{T_t\times d},
  \label{eq:tqgf_inputs}
\end{equation}
where $\mathbf{w}$ are the $T_t$ text tokens the LLM has already generated for its response, and $\mathbf{H}_\text{llm}\in\mathbb{R}^{T_t\times 3584}$ are the LLM's last-layer hidden states at those same positions. The conditioning stream $\mathbf{Z}$ supplies both the keys and the values of the block. The MLP widens each hidden state to $4d$ dimensions, and the reshape reads that vector as four consecutive $d$-dimensional positions, so $\mathbf{Z}$ is four times as long as the text sequence. The query stream, by contrast, looks the same tokens up in the Speech Generator's own embedding table $\mathbf{E}_\text{sg}$, giving one un-contextualised vector per token.

Each of the two TQGF layers then injects $\mathbf{Z}$ into the query stream. Writing $\tilde{\mathbf{Q}}$ and $\tilde{\mathbf{Z}}$ for the layer-normalised inputs, a layer computes
\begin{align}
  \mathbf{C} &= \Big(\sigma(\tilde{\mathbf{Q}}\mathbf{W}_g)\odot
    \mathrm{Attn}\big(\tilde{\mathbf{Q}}\mathbf{W}_q,\,\tilde{\mathbf{Z}}\mathbf{W}_k,\,\tilde{\mathbf{Z}}\mathbf{W}_v\big)\Big)\mathbf{W}_o,
    \label{eq:tqgf_gate}\\
  \mathbf{Q}' &= \mathbf{Q}+\mathbf{C},\qquad
  \mathbf{Q}'' = \mathbf{Q}'+\mathrm{FFN}(\mathbf{Q}'),
    \label{eq:tqgf_res}
\end{align}
where $\mathrm{Attn}$ is multi-head scaled dot-product attention with $h$ heads and no output projection, so $\mathrm{Attn}(\cdot)\in\mathbb{R}^{T_t\times h d_h}$; $\mathrm{FFN}$ is a position-wise feed-forward network with pre-norm and inner width $4d$, $\sigma$ is the logistic sigmoid and $\odot$ is element-wise multiplication. The projections satisfy $\mathbf{W}_q,\mathbf{W}_k,\mathbf{W}_v,\mathbf{W}_g\in\mathbb{R}^{d\times h d_h}$ and $\mathbf{W}_o\in\mathbb{R}^{h d_h\times d}$. We use $h=16$ heads of dimension $d_h=256$, so the gate acts in a $4096$-dimensional space rather than in the $896$-dimensional working space.

Plain cross-attention would send $\mathrm{Attn}(\cdot)\mathbf{W}_o$ straight into the residual stream; the factor $\sigma(\tilde{\mathbf{Q}}\mathbf{W}_g)$ in Equation~\eqref{eq:tqgf_gate} filters it first. The query holds one un-contextualised embedding per token, whereas the keys and values carry the LLM's semantic plan for the utterance, its prosodic intent and the dialogue history, along with information the speech head does not need. Gating before $\mathbf{W}_o$ lets each query token decide, coordinate by coordinate, how much of that signal to admit. Because Equation~\eqref{eq:tqgf_res} adds the gated branch to $\mathbf{Q}$ instead of replacing it, a closed gate falls back on the un-contextualised query rather than producing a zero-information state.

\subsection{VQ-VAE Body Representation}

Following LOM~\cite{lom2024}, we operate on 6D-rotation SMPL-X~\cite{smplx} parameters for body and hand articulation, together with FLAME~\cite{flame2017} expression coefficients for the face (the same face parameterisation that is embedded inside the SMPL-X face model). The per-frame body state is decomposed into four parts as summarized in Table~\ref{tab:body_dims}.

\bigskip\noindent
\begin{minipage}{\linewidth}
\centering\small
\captionof{table}{SMPL-X + FLAME feature decomposition used by the four LOM VQ-VAEs. The face branch combines the SMPL-X jaw rotation with FLAME~\cite{flame2017} expression coefficients; the remaining three branches are SMPL-X joint rotations (6D rotation representation).}
\label{tab:body_dims}
\setlength{\tabcolsep}{6pt}
\begin{tabular}{lrl}
\toprule
\textbf{Part} & \textbf{Dim} & \textbf{Composition} \\
\midrule
Face  & $106$ & jaw ($1 \times 6$) $+$ FLAME expression ($100$) \\
Hand  & $180$ & $30$ hand joints $\times\, 6$ \\
Upper & $78$  & $13$ upper-body joints $\times\, 6$ \\
Lower & $61$  & $9$ lower-body joints $\times\, 6$ $+$ translation ($3$) $+$ foot contact ($4$) \\
\bottomrule
\end{tabular}
\end{minipage}
\bigskip

The four per-part VQ-VAEs share codebook size $|\mathcal{C}|=256$ and are taken verbatim from the LOM \texttt{emage\_vq} release (architecture \texttt{VQVAEConvZero}, 30\,Hz, no temporal downsampling). We do not modify or fine-tune them.

\subsection{Motion Generator Hyperparameters}

\bigskip\noindent
\begin{minipage}{\linewidth}
\centering
\captionof{table}{Motion Generator architectural hyperparameters.}
\small
\begin{tabular}{ll}
\toprule
\textbf{Parameter} & \textbf{Value} \\
\midrule
Working dimension $d_m$               & 512 \\
Number of attention heads $h$         & 8 \\
Per-head dimension $d_h$              & 64 \\
TQGF depth $L_\text{TQGF}$            & 2 \\
Self-attention depth $L_\text{self}$  & 6 \\
TQGF FFN multiplier                   & 4 \\
Self-attention FFN multiplier         & 2 \\
Dropout                               & 0.0 \\
Periodic RoPE period $\mathcal{T}$    & 30 frames \\
RoPE cache length                     & 10{,}000 \\
Max motion length $T_m^{\max}$        & 600 frames (20\,s) \\
Per-part codebook size $|\mathcal{C}|$ & 256 \\
Motion frame rate                     & 30\,Hz \\
Speech-token rate                     & 12.5\,Hz \\
Ratio $T_m / T_s$                      & 2.4 \\
\bottomrule
\end{tabular}
\end{minipage}
\bigskip

\subsection{Stage-wise Training Hyperparameters}
\label{sec:supp_training_hparams}

\newcommand{\lr}[2]{$#1\!\times\!10^{#2}$}   %

\bigskip\noindent
\begin{minipage}{\linewidth}
\centering\small
\captionof{table}{Training configuration of the four stages (rows are settings; columns are stages). $\dagger$: Stage~4 starts from the best Stage~3d checkpoint. ``MG'': Motion Generator; ``SG'': Speech Generator; ``SP'': Speech Projector; ``LR'' is the peak learning rate. Warmup is given as a fraction of total steps unless noted.}
\label{tab:training_hparams}
\setlength{\tabcolsep}{4pt}
\begin{tabular*}{\linewidth}{@{\extracolsep{\fill}}lcccc@{}}
\toprule
\textbf{Setting} & \textbf{Stage 1} & \textbf{Stage 2} & \textbf{Stage 3 (a--d)} & \textbf{Stage 4}$^{\dagger}$ \\
\midrule
Trainable modules & SP & SG & SG + MG & LLM + SP + SG + MG \\
\addlinespace[2pt]
LR (SP)  & \lr{1}{-3} & --         & --         & \lr{1}{-6} \\
LR (SG)  & --         & \lr{1}{-4} & \lr{1}{-5} & \lr{5}{-7} \\
LR (MG)  & --         & --         & \lr{2}{-4} & \lr{2}{-6} \\
LR (LLM) & --         & --         & --         & \lr{2}{-7} \\
\addlinespace[2pt]
LR schedule & cosine & cosine & const + warmup & const + warmup \\
Warmup      & $0.30$ & $0.10$ & $0.03$         & $3000$ steps \\
Grad clip   & $1.0$  & $1.0$  & $1.0$          & $0.5$ + NaN guard \\
\addlinespace[2pt]
Effective batch size & $128$ & $128$ & $128$ & $128$ \\
Epochs & $1$ & $3$ & $\sim$5 per substage & until val.\ plateau \\
\bottomrule
\end{tabular*}
\end{minipage}
\bigskip

\subsection{Implementation Details}

\paragraph{Hardware and software.}
All training is conducted on $4$ GPUs using DeepSpeed ZeRO-2~\cite{deepspeed}, BF16 mixed precision and gradient checkpointing. The effective batch size is $128$ across all stages.

\paragraph{Model parameters.}
The total parameter count of Motion-Omni-Q7 is approximately $8.3$B, broken down as follows:
\begin{itemize}[leftmargin=*]
\itemsep0em
\item LLM Backbone (Qwen2.5-7B-Instruct): ${\sim}7.6$B
\item Speech Generator (Qwen2.5-0.5B-Instruct): ${\sim}0.5$B
\item Motion Generator ($4$ part-specific decoders, $d{=}512$): ${\sim}150$M
\item Speech Projector (2-layer MLP with stride-5): ${\sim}5$M
\end{itemize}

\paragraph{Training cost.}
The full four-stage training of Motion-Omni-Q7 takes approximately $960$ GPU-hours in total.

\subsection{Loss Composition and Multi-Task Balancing}

Let $\mathcal{L}_{\text{LM}}$ denote the LLM next-token cross-entropy, $\mathcal{L}_{\text{sp}}$ the speech-unit label-smoothed cross-entropy, and $\mathcal{L}_{\text{mo},b}$ the per-part motion cross-entropy on body part $b\in\{\mathrm{face},\mathrm{hand},\mathrm{upper},\mathrm{lower}\}$. The per-batch loss is
\begin{align}
  \mathcal{L} &= \mathcal{L}_{\text{LM}} + \mathcal{L}_{\text{sp}}\cdot\mathbb{1}[\text{speech task}] \nonumber\\
  &\quad + \sum_{b} w_b\,\mathcal{L}_{\text{mo},b}\cdot\mathbb{1}[\text{motion task}],
\end{align}
with $(w_{\mathrm{face}}, w_{\mathrm{hand}}, w_{\mathrm{upper}}, w_{\mathrm{lower}})$ proportional to the feature dimensions of Table~\ref{tab:body_dims} (i.e.\ $(106, 180, 78, 61)/425$). The Speech Projector has no dedicated loss term; it is trained through $\mathcal{L}_{\text{LM}}$ via backpropagation when the input contains audio (the projected Whisper features feed into the LLM, and the LLM CE loss gradient flows back through the projector). The masks $\mathbb{1}[\cdot]$ select the samples whose task requires the corresponding supervision: ASR contributes only $\mathcal{L}_{\text{LM}}$; TTS, TTSM and S2SM contribute $\mathcal{L}_{\text{LM}} + \mathcal{L}_{\text{sp}}$; TTSM and S2SM additionally contribute $\sum_b w_b\mathcal{L}_{\text{mo},b}$; T2T contributes only $\mathcal{L}_{\text{LM}}$ on text-only samples. When logging training curves, we average each per-task loss only over batches that contain at least one sample of the corresponding task, so that zero-padding from other tasks does not artificially depress the reported loss.

\clearpage
\section{Data Pipeline Details}
\label{sec:data_details}

\subsection{ASR Audio Filtering}
\label{sec:asr_filtering}

For Stage~1 ASR supervision, we discard transcripts containing non-English characters, transcripts shorter than two words, and utterances outside the $1$--$30$\,s duration range.
For recordings, we require an energy-distribution SNR proxy of at least $15$\,dB.
Whisper-large-v3~\cite{whisper} then retranscribes every candidate, and only pairs with $\mathrm{WER}\leq 5\%$ are retained.
For all-uppercase transcripts, we use the Whisper-cased transcript only when the normalised word sequence matches exactly ($\mathrm{WER}=0$); otherwise the sample is removed.
Dataset-specific punctuation placeholders are normalised after filtering, and duplicate audio paths are removed.

\subsection{TTS Audio Filtering}
\label{sec:tts_filtering}

All Stage~2 TTS waveforms, including those derived from Emilia text and InstructS2S assistant-response text, are synthesised with Qwen3-TTS~\cite{qwen3tts}; the speech-unit targets use the GLM-4-Voice tokenizer and vocabulary~\cite{glm4voice}.
For Stage~2 TTS supervision, we first remove text containing non-English characters or outside the range of $2$--$200$ words, then require that the waveform is present and decodable, lasts $1$--$25$\,s.
Text--audio agreement is checked by Whisper-large-v3, retaining only samples with $\mathrm{WER}<5\%$.
We further apply acoustic event detection (AED) to each waveform and reject those whose non-speech portion exceeds $30\%$ of the total duration.

\subsection{Weighted VQ-VAE Reconstruction Error}
\label{sec:vqvae_error}

The weighted VQ-VAE reconstruction error $L_1^\text{vq}$ measures how faithfully the frozen LOM VQ-VAE codebook can represent a given motion sample. For each sample in the Stage~3 corpus, we compute this metric as follows.

\paragraph{Step~1: Feature extraction.}
The original motion parameters (stored as SMPL-X axis-angle rotations, FLAME expression coefficients and global translation) are converted into per-part continuous feature vectors following the body decomposition in Table~\ref{tab:body_dims}: face ($D_\text{face}=106$), hand ($D_\text{hand}=180$), upper body ($D_\text{upper}=78$) and lower body ($D_\text{lower}=61$). Axis-angle joint rotations are converted to 6D rotation representations before concatenation.

\paragraph{Step~2: Encode, quantise, decode.}
Each part's feature sequence $\hat{m}^{(b)}_{1:T} \in \mathbb{R}^{T \times D_b}$ is passed through the frozen LOM VQ-VAE for that part:
\[
  \hat{m}^{(b)} \;\xrightarrow{\text{encoder}}\; z^{(b)} \;\xrightarrow{\text{quantise}}\; c^{(b)} \;\xrightarrow{\text{decoder}}\; \tilde{m}^{(b)},
\]
where the quantisation step maps each latent vector to its nearest codebook entry ($|\mathcal{C}|=256$). The reconstructed sequence $\tilde{m}^{(b)}_{1:T}$ has the same dimensionality and frame rate as the original.

\paragraph{Step~3: Per-part L1 error.}
For each body part $b$, we compute the mean absolute error averaged over all frames and all feature dimensions:
\begin{equation}
  \ell_b \;=\; \frac{1}{T \cdot D_b} \sum_{t=1}^{T} \bigl\lVert \hat{m}^{(b)}_t - \tilde{m}^{(b)}_t \bigr\rVert_1.
\end{equation}

\paragraph{Step~4: Dimension-weighted aggregation.}
The final score is the dimension-weighted average across parts:
\begin{align}
  &L_1^\text{vq}(\hat{m}) = \frac{\sum_{b} D_b \cdot \ell_b}{\sum_{b} D_b} \nonumber\\
  &= \frac{106\,\ell_\text{face} + 180\,\ell_\text{hand} + 78\,\ell_\text{upper} + 61\,\ell_\text{lower}}{425},
\end{align}
so that each feature dimension contributes equally to the aggregate score regardless of which body part it belongs to.

\paragraph{Interpretation.}
$L_1^\text{vq}$ measures \emph{supervision fidelity}: it quantifies how faithfully the discrete VQ codes (which serve as training targets for the Motion Generator) represent the underlying continuous motion. A high $L_1^\text{vq}$ indicates that the quantisation step maps the latent to a distant codebook entry, so the code assigned by the teacher VQ-VAE is a poor proxy for the intended motion and training on such samples injects noisy label supervision. However, a low $L_1^\text{vq}$ alone does not guarantee the motion is useful as supervision: the codebook may faithfully represent a motion sequence that is nevertheless poorly aligned with the driving speech. The complementary beat correlation score (Section~\ref{sec:bc_pipe}) catches exactly this failure mode by measuring speech-motion rhythmic coupling. Together the two metrics form a dual filter: $L_1^\text{vq}$ ensures label fidelity (the codes reliably represent the motion), while BC ensures speech-motion coupling (the motion is rhythmically aligned with the audio). The curriculum (Section~\ref{sec:dataset}) leverages the combined score to expose the network to high-fidelity, well-aligned samples first.

\subsection{Beat Correlation Score (BC)}
\label{sec:bc_pipe}

Given the assistant audio $a$ and LOM-predicted motion $\hat{m}$, we first extract a body-level kinematic envelope $v_t = \lVert \dot{X}_{\text{upper},t}\rVert_2$ on the upper body at $30\,\mathrm{Hz}$, and detect kinematic beats $\mathcal{M} = \{t : v_t < v_{t-1},\, v_t < v_{t+1},\, v_t > 0.3\bar{v}\}$ as the local minima of $v$ above a relative threshold. We extract audio onsets $\mathcal{O}$ from $a$ via librosa~\cite{librosa} and compute the forward (audio$\to$motion) Gaussian-kernel beat-distance score
\begin{align}
  &\mathrm{BC}(a, \hat{m}) = \nonumber\\
  &\frac{1}{|\mathcal{O}|}\sum_{t_o\in\mathcal{O}}
  \exp\!\left(-\frac{\min_{t_m\in\mathcal{M}} (t_o-t_m)^2}{2\sigma^2}\right), \nonumber\\
  &\sigma = 9\text{ frames}.
\end{align}
This is the audio$\to$motion direction of the EMAGE BC formulation~\cite{emage2024}, restricted to a single body-level envelope. We use this established BC direction for curriculum scoring and the main descriptive comparison. The bidirectional $\mathrm{BC}_\text{bi}$ defined in Equations~\ref{eq:bc_am}--\ref{eq:bc_bi} is evaluated only as an exploratory diagnostic.

\paragraph{Bidirectional beat diagnostic.}
We additionally examine a bidirectional beat-correlation diagnostic, $\mathrm{BC}_\text{bi}$. Standard BC matches audio onsets $\mathcal{O}$ to nearby motion beats $\mathcal{M}$ in one direction. Since both audio and motion are generated in our setting, we average both matching directions:
\begin{align}
  \mathrm{BC}_{a\to m}
  &= \frac{1}{|\mathcal{O}|}\sum_{t_o\in\mathcal{O}}
    \exp\!\left(-\frac{\min_{t_m\in\mathcal{M}}(t_o-t_m)^2}{2\sigma^2}\right),
    \label{eq:bc_am}\\
  \mathrm{BC}_{m\to a}
  &= \frac{1}{|\mathcal{M}|}\sum_{t_m\in\mathcal{M}}
    \exp\!\left(-\frac{\min_{t_o\in\mathcal{O}}(t_o-t_m)^2}{2\sigma^2}\right),
    \label{eq:bc_ma}\\
  \mathrm{BC}_\text{bi}
  &= \tfrac{1}{2}\left(\mathrm{BC}_{a\to m}+\mathrm{BC}_{m\to a}\right),
    \label{eq:bc_bi}
\end{align}
where audio onsets are detected with \texttt{librosa.onset.onset\_detect}~\cite{librosa}, motion beats are local minima of the upper-body kinetic-energy envelope above a relative threshold, and $\sigma=0.3\,\mathrm{s}$ follows the LOM setup. The two directions respectively penalise unmatched audio onsets and unmatched motion beats.

We compare the per-pair ordering induced by $\mathrm{BC}_\text{bi}$ with the majority R1 rhythm verdict on the 100 human-evaluated pairs. Treating score differences below $0.07$ as ties gives $46\%$ agreement, against $40\%$ for standard BC under the same tie-aware protocol. On a set of this size neither figure separates from the chance reference, so we report $\mathrm{BC}_\text{bi}$ as an exploratory diagnostic rather than a validated perceptual metric.

\subsection{Robust Percentile Normalisation}

To form the combined quality score $s(x)$ of Equation~(1) of the main paper we robustly normalise each metric: the 5th percentile is mapped to $0$, the 95th percentile to $1$, and values outside the $[0,1]$ range are clipped. The $L_1^\text{vq}$ axis is additionally inverted (large~$\to$~0, small~$\to$~1) so that $1$ denotes the highest quality in both axes. Samples for which either metric is unavailable (e.g.\ silent audio, degenerate motion) are excluded from the ranking and do not participate in the Stage~3 curriculum.

\section{Rendering Pipeline}
\label{sec:supp_render}

To produce the video stimuli used in the human preference study, we render each motion sequence as a 30\,fps video with synchronised response audio. The rendering pipeline uses the following assets and tools:

\begin{itemize}[leftmargin=*]
\itemsep0em
 \item \textbf{Mesh template}: Meshcapade SMPL-X female mesh (\texttt{SMPLX-female.obj}, CC-BY-4.0 licence), which provides a canonical UV layout with 42{,}064 UV-seam-duplicated vertices mapped to 10{,}475 unique SMPL-X vertices.
 \item \textbf{Material}: PBR metallic-roughness material with an albedo map (\texttt{f\_01\_alb.002.palmfix.png}, palm-tone-corrected) and a normal map (\texttt{f\_01\_nrm.002.png}) from the Meshcapade texture samples; metallic factor $0.0$, roughness factor $0.6$.
 \item \textbf{Lighting}: three-point directional lights (key, fill, rim) plus an ambient term estimated from the median of an HDRI environment map (\texttt{studio\_small\_03\_4k.hdr}).
 \item \textbf{Renderer}: \texttt{pyrender} with headless EGL backend (\texttt{PYOPENGL\_PLATFORM=egl}), producing $720\times720$ frames.
 \item \textbf{Body model}: SMPL-X~\cite{smplx} with FLAME~\cite{flame2017} expression blendshapes, driven by the decoded per-frame 6D-rotation parameters and expression coefficients from the LOM VQ-VAE output.
\end{itemize}

Frames are composited into an MP4 (H.264) with the response audio muxed in, using FFmpeg. The same rendering configuration is applied identically to all systems (Motion-Omni-Q7 and cascaded baselines) so that visual differences arise solely from the motion stream.

\section{Human Evaluation Protocol}
\label{sec:supp_human_pref}

Because automatic gesture metrics (FGD, BC) are imperfect proxies for perceived motion quality, we additionally conduct a paired A/B/tie human preference study.

\paragraph{Evaluation set.}
We select 25 clips per comparison arm from the external SwDA-500 evaluation set, covering both short ($4$--$10\,\mathrm{s}$) and long ($\geq 10\,\mathrm{s}$) responses across the sampled dialogue topics. Clips with degenerate responses ($< 40$ characters) or missing motion files are excluded.

\paragraph{Study design.}
Each of the 25 clips is compared across four arms, yielding 100 total pairs: B1 is Qwen2.5-Omni + LOM, B2 is Qwen2.5-Omni + EMAGE, B3 is MO-audio + LOM, and B4 is MO-audio + EMAGE. Arms B1/B2 use a different speech model, so both speech and motion differ from Motion-Omni-Q7; arms B3/B4 share Motion-Omni-Q7's own audio, isolating the motion contribution.

\paragraph{Interface and rating.}
Each pair is presented as two side-by-side rendered videos (left = System~A, right = System~B) with the system assignment randomised per pair (fixed seed $= 42$). Annotators judge each pair on \textbf{three motion-related dimensions} independently, choosing one of \emph{A is better}, \emph{B is better}, or \emph{Tie} per dimension.

\paragraph{Annotators.}
Four annotators independently rate all 100 pairs on dimensions R1--R3 (video+audio). Per-pair verdicts are determined by majority vote across the four raters. Inter-annotator agreement is measured by mean pairwise Cohen's kappa.

\subsection{Annotation Rubric}
\label{sec:supp_rubric}

The following rubric is provided to all annotators.

\paragraph{R1 -- Rhythm.} Does the body motion match the rhythm of the speech? We instruct annotators to check whether hand gestures or head movements accompany stressed words, whether the body pauses when the speech pauses (rather than continuing to gesture), and whether gesture density tracks speech rate. A candidate is preferred when it feels like ``this person is actually saying these words.''

\paragraph{R2 -- Semantic Alignment.} Do the gestures convey the \emph{content} being spoken? Positive examples include counting fingers for enumerated items, pointing for spatial references, tracing shapes for described objects, and head nods/shakes for affirmation/negation. Generic rhythmic hand-waving (e.g.\ open-palm emphasis) does \emph{not} count as semantic alignment. If neither side shows clear semantic gestures, annotators select tie.

\paragraph{R3 -- Body Naturalness.} Does the motion look like a real person in conversation? Annotators check for (a)~mechanical artefacts (jitter, interpenetration, sudden joint pops, foot sliding), (b)~posture issues (frozen pose, exaggerated public-speaking stance), and (c)~overall relaxation (subtle weight shifts, natural idle motion). We emphasise that large motion amplitude does not imply naturalness.

\paragraph{Scoring principles.}
Annotators are instructed to (1)~score each dimension independently without first forming an overall impression, (2)~treat tie as a legitimate option.

\section{Spoken Dialogue Quality on VoiceBench}
\label{sec:supp_voicebench}

Public SDM benchmarks~\cite{voicebench,DBLP:conf/emnlp/MaTG25} evaluate dialogue capabilities.
Table~\ref{tab:voicebench} gives the per-subset VoiceBench~\cite{voicebench} scores summarised in Section~\ref{sec:voicebench}. The listed systems differ by orders of magnitude in the amount of speech seen during pretraining and in whether speech generation is their primary objective, so we report the scores as context for the capability-retention check rather than as a like-for-like ranking, and do not mark best values.

\begin{table}[H]
\centering\small
\setlength{\tabcolsep}{4pt}
\caption{Spoken dialogue quality on VoiceBench~\cite{voicebench}. Higher is better throughout. GPT-4o-Audio is served through an API and its weights are not released; every other system releases its weights. Dashes denote scores not reported.}
\label{tab:voicebench}
\resizebox{\textwidth}{!}{%
\begin{tabular}{lcccccccccc}
\toprule
\textbf{Model} & \textbf{Alpaca} & \textbf{Common} & \textbf{Wild} & \textbf{SD-QA} & \textbf{MMSU} & \textbf{OBQA} & \textbf{BBH} & \textbf{IFEval} & \textbf{AdvBench} & \textbf{Overall} \\
\midrule
\multicolumn{11}{l}{\emph{Closed-weight (API only)}} \\
GPT-4o-Audio~\cite{gpt4oaudio}    & 4.78 & 4.49 & 4.58 & 75.50 & 80.25 & 89.23 & 84.10 & 76.02 & 98.65 & 86.75 \\
\midrule
\multicolumn{11}{l}{\emph{Open-weight}} \\
Kimi-Audio~\cite{kimiaudio}       & 4.46 & 3.97 & 4.20 & 63.12 & 62.17 & 83.52 & 69.70 & 61.10 & 100.00 & 76.91 \\
Qwen2.5-Omni~\cite{qwen25omni}    & 4.49 & 3.93 & --   & 55.71 & 61.32 & 81.10 & --    & 52.87 & 99.42 & --    \\
VITA-1.5~\cite{vita15}            & 4.21 & 3.66 & 3.48 & 38.88 & 52.15 & 71.65 & 55.30 & 38.14 & 97.69 & 64.53 \\
LLaMA-Omni~\cite{llamaomni2024}   & 3.70 & 3.46 & 2.92 & 39.69 & 25.93 & 27.47 & 49.20 & 14.87 & 11.35 & 41.12 \\
Ex-Omni~\cite{exomni2025}         & 2.57 & 2.87 & 2.19 & 40.14 & 24.24 & 25.49 & 50.40 & 16.22 & 83.08 & 43.57 \\
Moshi~\cite{moshi}                & 2.01 & 1.60 & 1.30 & 15.64 & 24.04 & 25.93 & 47.40 & 10.12 & 44.23 & 29.51 \\
Mini-Omni2~\cite{DBLP:journals/corr/abs-2410-11190}       & 2.32 & 2.18 & 1.79 & 9.31  & 24.27 & 26.59 & 46.40 & 11.56 & 57.50 & 33.49 \\
\midrule
Motion-Omni-Q7 (ours)             & 3.17 & 3.35 & 2.97 & 39.78 & 27.36 & 26.59 & 51.00 & 20.87 & 73.27 & 47.63 \\
\bottomrule
\end{tabular}%
}
\end{table}

\section{Speech Naturalness Proxy}
\label{sec:supp_utmos}

Because WER measures only intelligibility, we additionally report UTMOSv2~\cite{utmosv2} on SwDA-500 generated audio as an automatic proxy for speech naturalness. UTMOSv2 is a learned predictor of mean opinion score and does not replace a listening test; we report it as a lightweight speech-quality check under conversational wording rather than as a competitive comparison, and the reference systems below are dedicated TTS models that synthesise a supplied sentence rather than plan a dialogue response.

\begin{table}[H]
\centering\small
\setlength{\tabcolsep}{5pt}
\caption{Automatic speech naturalness proxy on SwDA-500 generated audio. In each column, \textbf{bold} marks the best value and \underline{underline} the second best.}
\label{tab:utmos}
\begin{tabular}{lc}
\toprule
\textbf{Model / System} & \textbf{UTMOSv2 en mean}$\uparrow$ \\
\midrule
Qwen3-TTS~\cite{qwen3tts} & \textbf{4.05} \\
CosyVoice 3~\cite{cosyvoice3} & \underline{3.92} \\
GLM-TTS~\cite{DBLP:journals/corr/abs-2512-14291} & 3.31 \\
VoxCPM1.5~\cite{DBLP:journals/corr/abs-2509-24650} & 3.17 \\
F5-TTS~\cite{f5tts} & 3.15 \\
CosyVoice~\cite{DBLP:journals/corr/abs-2407-05407} & 3.06 \\
\midrule
Motion-Omni-Q7 (ours) & 3.77 \\
\bottomrule
\end{tabular}
\end{table}

Motion-Omni-Q7 scores $3.77$, above GLM-TTS, VoxCPM1.5, F5-TTS and CosyVoice, and below Qwen3-TTS ($4.05$) and CosyVoice~3 ($3.92$).

\section{Fusion and Curriculum Ablations}
\label{sec:supp_ablations}

\subsection{Motion-Conditioning Fusion}

To isolate the fusion design, all variants start from the same Stage~2 checkpoint and use the same Stage~3a data and optimisation schedule. The only changed factor is how the motion query incorporates contextual information. TQGF/LLM K/V reads keys and values directly from the LLM; the remaining variants read Speech Generator states. We report the established unidirectional BC metric only.

\begin{table}[H]
\centering\small
\caption{Motion-conditioning ablation under the Stage~3a setting.}
\label{tab:tqgf_ablation}
\begin{tabular}{clc}
\toprule
\textbf{No.} & \textbf{Fusion / context source} & \textbf{BC}$\uparrow$ \\
\midrule
1 & TQGF / LLM K/V              & 0.827 \\
2 & Concatenation / Speech Generator K/V & 0.809 \\
3 & GLU / Speech Generator K/V  & 0.799 \\
4 & Cross-attention / Speech Generator K/V & 0.905 \\
5 & FiLM / Speech Generator K/V & 0.852 \\
6 & TQGF / Speech Generator K/V & \textbf{0.912} \\
\bottomrule
\end{tabular}
\end{table}

The comparison supports two scoped observations: Speech Generator context is more useful than direct LLM context for the same TQGF operator (No.~6 vs.\ No.~1), and gated fusion is modestly better than plain cross-attention under this fixed Stage~3a setup (No.~6 vs.\ No.~4). It does not isolate every internal component of TQGF.

\subsection{Stage-wise Teacher-Reference FGD}

Using the same LOM latent encoder and teacher-reference definition as the main motion evaluation, FGD decreases through all curriculum stages and remains lowest after Stage~4 joint fine-tuning.

\begin{table}[H]
\centering\small
\caption{Teacher-reference FGD across training stages.}
\label{tab:curriculum_fgd}
\begin{tabular}{lc}
\toprule
\textbf{Checkpoint} & \textbf{FGD}$\downarrow$ \\
\midrule
Stage~3a & 0.3974 \\
Stage~3b & 0.3258 \\
Stage~3c & 0.3079 \\
Stage~3d & 0.3075 \\
Stage~4  & \textbf{0.3040} \\
\bottomrule
\end{tabular}
\end{table}

These results are consistent with progressive improvement under the curriculum.

\subsection{Quality--Latency Trade-off}
\label{sec:supp_quality_latency}

\begin{figure}[H]
\centering
\begin{tikzpicture}
\begin{axis}[
  width=0.78\textwidth, height=5.4cm,
  xmode=log, log basis x=10,
  xlabel={Real-time factor RTF (lower is better)},
  ylabel={Beat correlation BC ($\times10^{-1}$)},
  xmin=0.45, xmax=45, ymin=7.18, ymax=7.86,
  ytick={7.2,7.4,7.6,7.8},
  xtick={1,10}, xticklabels={$1$,$10$},
  grid=major, grid style={dashed, gray!22},
  label style={font=\small}, tick label style={font=\small},
  axis line style={gray!60},
]
\draw[dashed, gray!70] (axis cs:1,7.18) -- (axis cs:1,7.86);
\addplot[only marks, mark=*, mark size=2.1pt, draw=black!65, fill=black!25] coordinates {
  (18.34,7.56) (8.79,7.31) (4.39,7.67) (0.84,7.32)
};
\addplot[only marks, mark=star, mark size=4.4pt, draw=red!75!black, thick] coordinates {(0.78,7.59)};
\node[font=\footnotesize, anchor=east] at (axis cs:17.0,7.56) {Qwen2.5-Omni + LOM};
\node[font=\footnotesize, anchor=south] at (axis cs:8.79,7.335) {Qwen2.5-Omni + EMAGE};
\node[font=\footnotesize, anchor=south] at (axis cs:4.39,7.695) {MO-audio + LOM};
\node[font=\footnotesize, anchor=north] at (axis cs:0.95,7.30) {MO-audio + EMAGE};
\node[font=\footnotesize, anchor=south west, red!75!black] at (axis cs:0.72,7.615) {\textbf{Motion-Omni-Q7}};
\end{axis}
\end{tikzpicture}
\caption{Motion quality against response latency on SwDA-500. Circles are two-stage cascades; the star is our end-to-end model. Points closer to the top-left are better. The dashed line marks $\mathrm{RTF}=1$, below which a system responds faster than real time.}
\label{fig:quality_latency}
\end{figure}
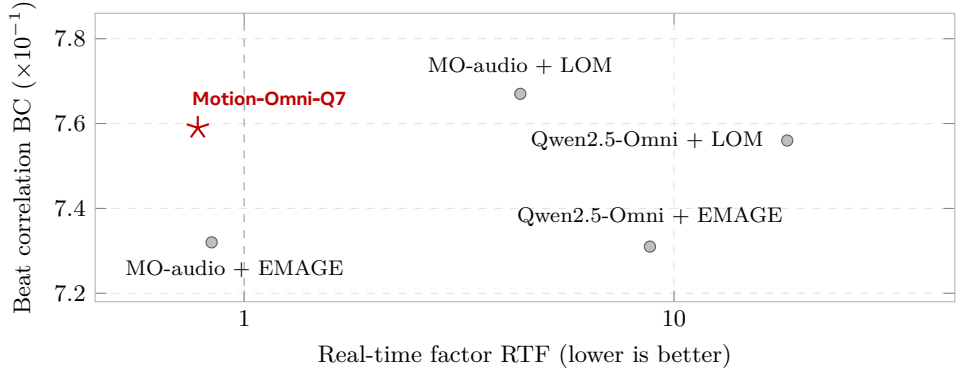

Figure~\ref{fig:quality_latency} jointly plots the BC values from Table~\ref{tab:motion} and the RTF values from Table~\ref{tab:latency}.
Motion-Omni-Q7 is the only evaluated system that combines faster-than-real-time generation ($\mathrm{RTF}<1$) with $\mathrm{BC}>7.5$.
MO-audio + EMAGE has similar latency ($0.84$ versus $0.78$) but lower BC ($7.32$ versus $7.59$), whereas MO-audio + LOM attains higher BC ($7.67$) at an RTF of $4.39$ and $5.4{\times}$ the response time.

\section{Negative Result: Video LLM-as-Judge Calibration}
\label{sec:supp_videojudge}

We initially intended to scale the rubric scoring of Section~\ref{sec:supp_human_pref} with a video-input LLM-as-judge (Gemini~3.1~Pro~\cite{gemini25}). After five rounds of prompt iteration on a 26-clip multi-turn calibration set and one separate pairwise-ranking experiment, we concluded that the judge is not a viable replacement for human ratings on this task. We document the result here as a negative finding; the calibration artefacts (annotation files, IAA reports, and prompts) are included in our planned release.

\paragraph{Calibration set and protocol.}
We collected $26$ multi-turn rendered clips spanning the same systems (Motion-Omni-Q7, B1, B2), scored them once with a single trained human annotator on a broader pilot rubric and once with Gemini~3.1~Pro, and iterated the judge prompt over five revisions (\texttt{iter1}--\texttt{iter5}). The revisions tightened anchors for speech quality, conversational body motion, semantic alignment, and multi-turn continuation. For each iteration we report the per-dimension Pearson $r$ between the single human and the judge, and the count of \emph{large disagreements} ($|G - H| \geq 2$ in the $1$--$5$ scale).

\paragraph{Absolute scoring is not viable.}
The fifth and best iteration achieves $r=+0.32$ on speech quality, $r=+0.18$ on R3 (Body Naturalness), and $r=+0.17$ on multi-turn coherence, while the remaining pilot dimensions, including R1 and R2, have $r\leq 0$. The total number of large disagreements drops from $21$ at iter1 to $10$ at iter5, but the average Pearson $r$ across the pilot rubric only reaches $+0.05$, an order of magnitude below the $0.50$--$0.70$ band reported for LLM-as-judge on text dialogue benchmarks~\cite{llamaomni2024} and below the $0.30$--$0.50$ band typical of subjective video-and-motion evaluations. Three independent root causes contribute: (i)~the $26$-clip $\times$ $1$--$5$ integer scale gives a per-dimension $r$ confidence interval of roughly $\pm 0.4$ which prompt iteration cannot escape; (ii)~the judge model has hard ceilings, including indistinguishable lip-sync from preview-quality video and a systematic R3 over-rating of large-amplitude motion that human raters perceive as over-presented; (iii)~R2 has a real disagreement among human raters about whether emphasis-time gestures count as semantic alignment, which no judge prompt can resolve.

\paragraph{Pairwise ranking is worse, not better, in our setting.}
Following the LMSYS Arena recipe, we additionally implemented a pairwise judge that, given two same-audio clips with different motion streams, picks the preferred one. On $13$ pairs the judge agrees with the human's aggregate pilot-rubric ranking on $4/13 = 31\%$, with Cohen's $\kappa = 0.079$, while a simpler ``score each clip in absolute terms and pick the higher-scoring one'' baseline reaches $5/13 = 38\%$. Because the two clips share the same audio, non-motion criteria are forced to ties and the pairwise verdict is dominated by R3 (Body Naturalness), which is exactly the dimension on which the judge has the largest systematic bias. Pairwise is therefore not a fix for the absolute-scoring failure: it amplifies the judge's weakest signal rather than averaging it out.

\paragraph{Decision.}
Given that only speech quality reaches a usable correlation on absolute scoring, that pairwise ranking does not recover the motion dimensions, and that the human-evaluation pipeline (Section~\ref{sec:supp_human_pref}) was already operational, we removed the video LLM-as-judge from the main evaluation entirely; the headline numbers in the main paper rely on human ratings only. The final protocol retains only R1--R3 and replaces per-clip $1$--$5$ scoring with the paired A/B/tie comparison described in Section~\ref{sec:supp_human_pref}. We plan to include the calibration data, prompts, and IAA tooling in the release; in our setting, an off-the-shelf multimodal LLM is not yet a substitute for human annotation of rendered conversational-digital-human clips.

\section{Qualitative Visualisations}
\label{sec:qualitative}

\FloatBarrier
\renewcommand{\topfraction}{0.92}
\renewcommand{\bottomfraction}{0.9}
\renewcommand{\textfraction}{0.06}
\renewcommand{\floatpagefraction}{0.8}
This section shows rendered frames from the released demonstration clips. All frames come from the rendering pipeline of Section~\ref{sec:supp_render}; the full videos are part of uploaded files.

\begin{figure}[!ht]
\centering
\begin{subfigure}{\textwidth}
\centering
\includegraphics[width=\textwidth]{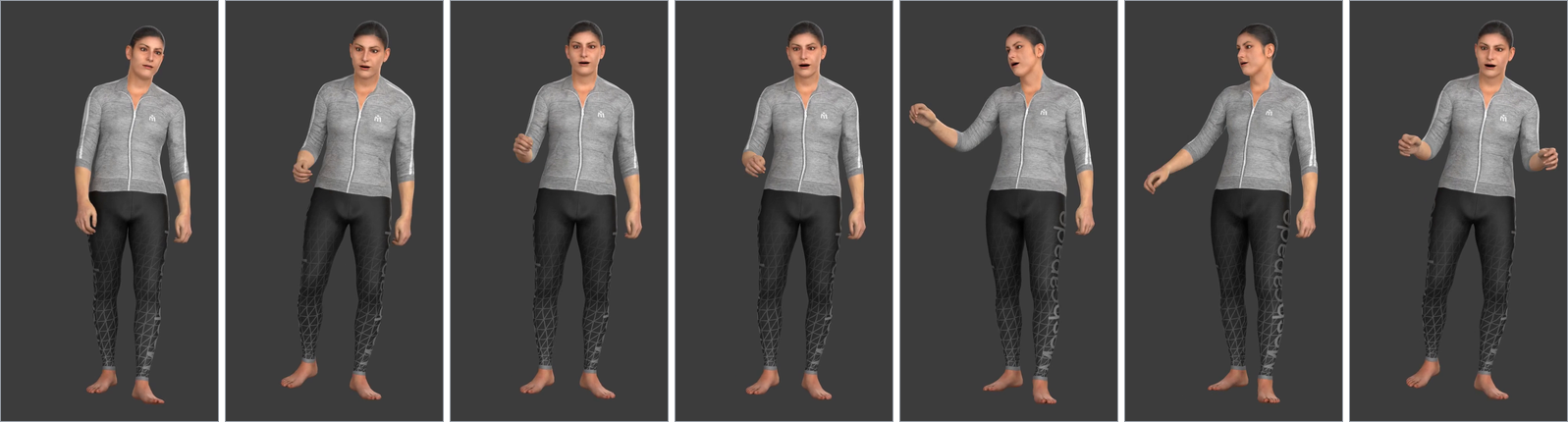}
\caption{``Here are five positive character traits: honesty, empathy, kindness, confidence, and curiosity.'' Frames at $t=1.1,\,1.9,\,2.6,\,3.9,\,4.8,\,6.0,\,6.9$\,s.}
\end{subfigure}
\vspace{4pt}
\begin{subfigure}{\textwidth}
\centering
\includegraphics[width=\textwidth]{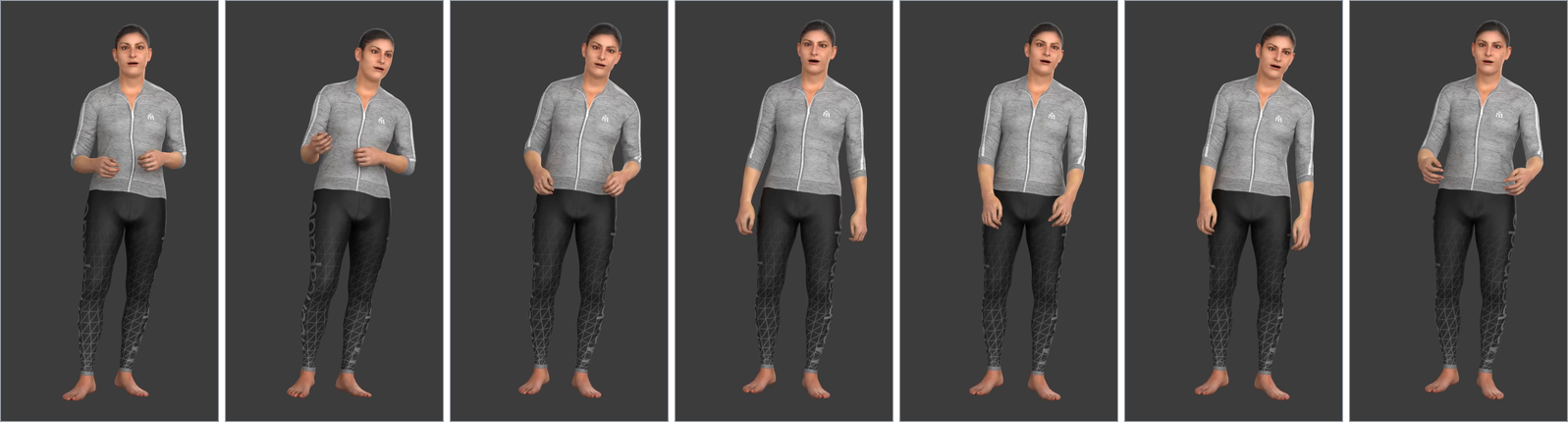}
\caption{``Supervisors can provide constructive feedback by being specific, timely, and respectful, focusing on behavior rather than personality.'' Frames at $t=1.4,\,2.3,\,3.4,\,4.3,\,5.2,\,6.6,\,8.0$\,s.}
\end{subfigure}
\vspace{4pt}
\begin{subfigure}{\textwidth}
\centering
\includegraphics[width=\textwidth]{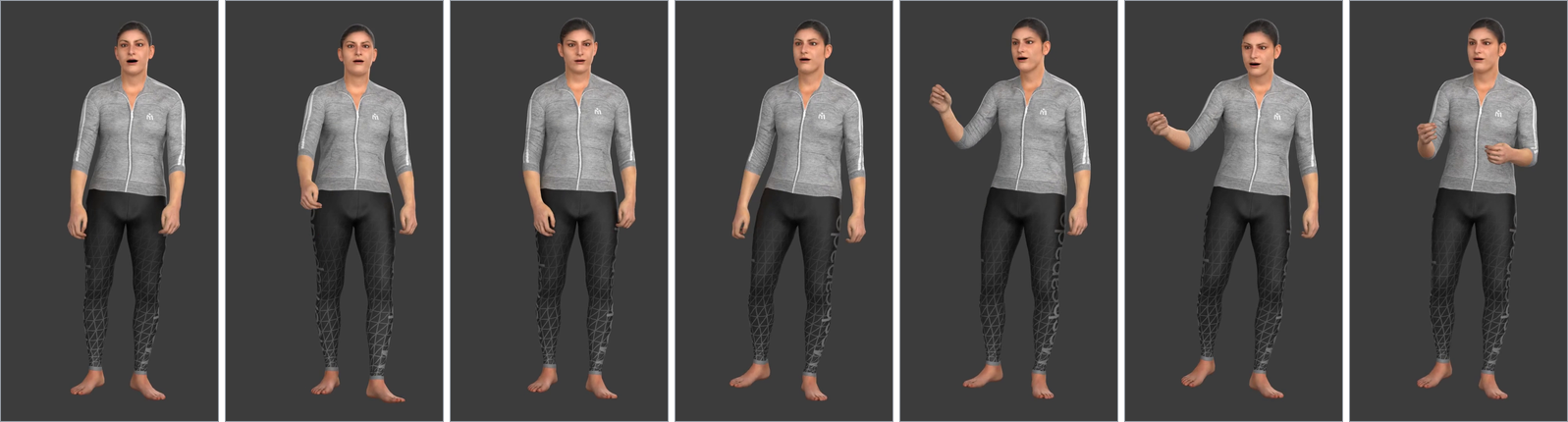}
\caption{``I think the show is trying to convey that social relationships are complex and can be both positive and negative, and that they can have a significant impact on our mental health and well-being.'' Frames at $t=1.2,\,2.3,\,4.2,\,5.2,\,7.0,\,7.7,\,8.5$\,s.}
\end{subfigure}
\caption{Motion-Omni-Q7 outputs on three released clips. Each row shows seven frames of a single response in temporal order, together with the text the model spoke in that response; the speech and the motion are produced in the same autoregressive pass. Frames are the seven mutually most distinct upper-body poses subject to a minimum temporal gap, not uniform samples.}
\Description{Three rows of seven rendered frames each, showing Motion-Omni-generated full-body motion for three spoken responses.}
\label{fig:qual_timeline}
\end{figure}

\begin{figure}[p]
\centering
\setlength{\parskip}{0pt}
\begin{subfigure}{\textwidth}
\centering
\includegraphics[width=0.9\textwidth]{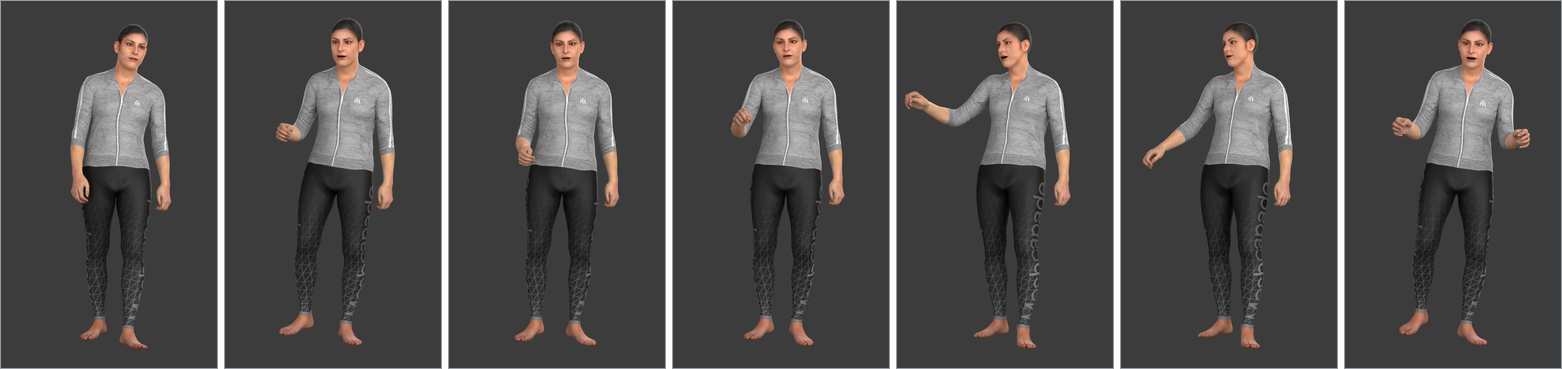}
\caption{Motion-Omni-Q7 (ours).}
\end{subfigure}
\vspace{3pt}
\begin{subfigure}{\textwidth}
\centering
\includegraphics[width=0.9\textwidth]{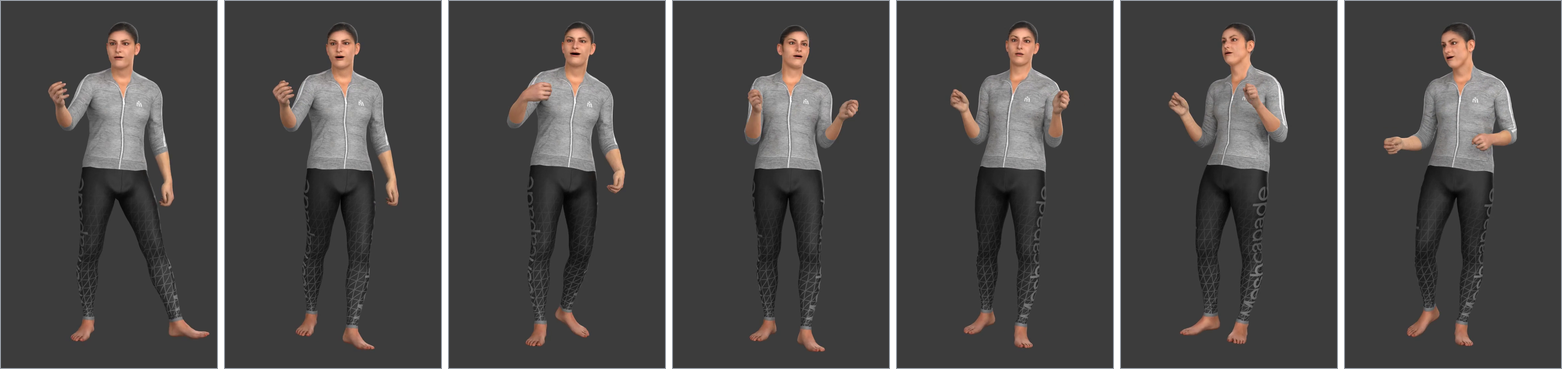}
\caption{MO-audio + LOM.}
\end{subfigure}
\vspace{3pt}
\begin{subfigure}{\textwidth}
\centering
\includegraphics[width=0.9\textwidth]{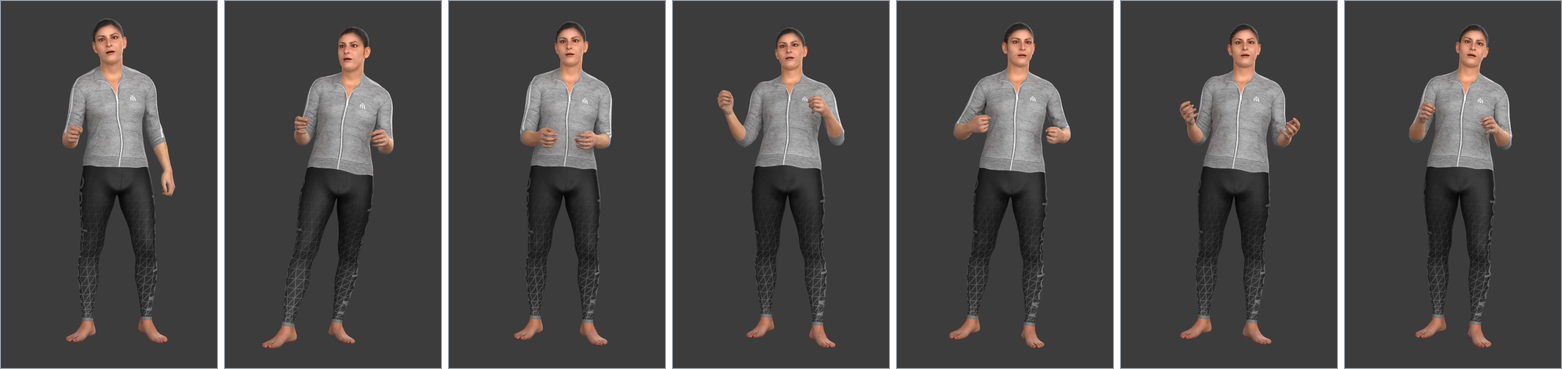}
\caption{MO-audio + GestureLSM.}
\end{subfigure}
\vspace{3pt}
\begin{subfigure}{\textwidth}
\centering
\includegraphics[width=0.9\textwidth]{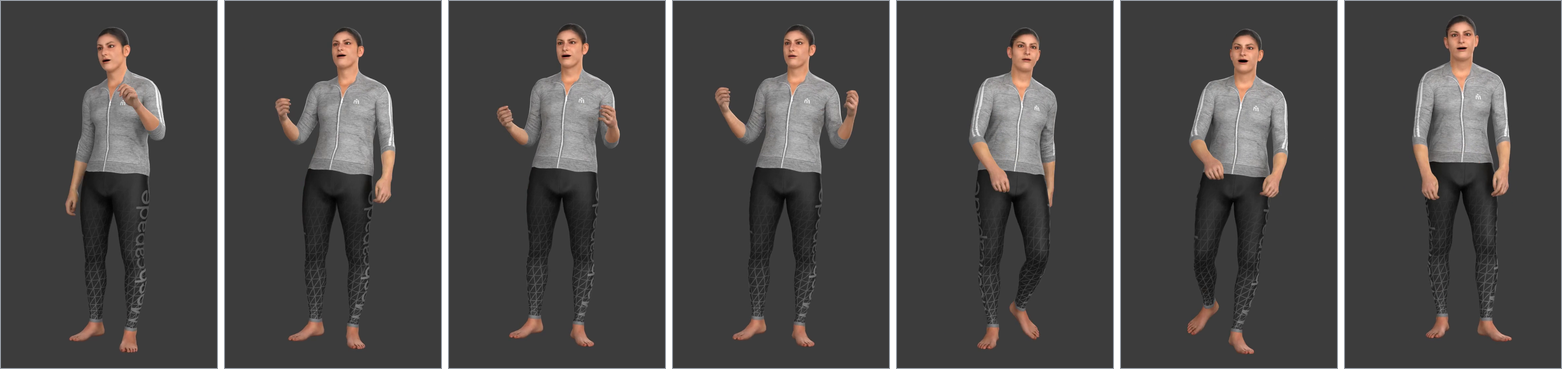}
\caption{MO-audio + EMAGE.}
\end{subfigure}
\vspace{3pt}
\begin{subfigure}{\textwidth}
\centering
\includegraphics[width=0.9\textwidth]{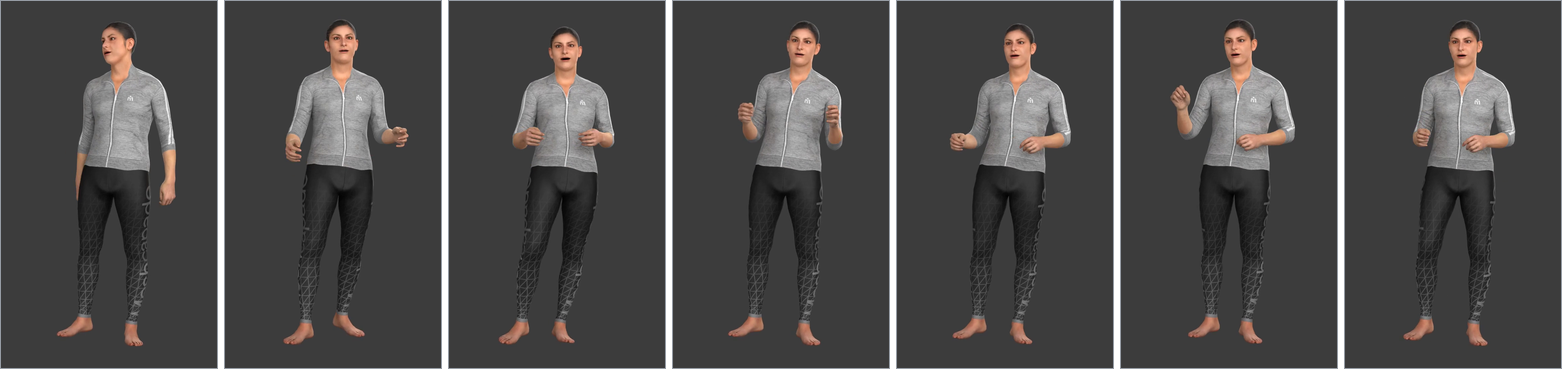}
\caption{MO-audio + MambaTalk.}
\end{subfigure}
\caption{The same response audio driving five different motion sources. All rows use the speech Motion-Omni-Q7 generated for ``Here are five positive character traits: honesty, empathy, kindness, confidence, and curiosity.'', sampled at the same seven timestamps ($t=1.1$ to $6.9$\,s in steps of about $0.98$\,s) and rendered with an identical pipeline, so the only variable across rows is the motion stream.}
\Description{Five rows of seven rendered frames each, one row per motion generation system, all driven by the same speech audio at the same timestamps.}
\label{fig:qual_systems}
\end{figure}

\paragraph{Frame selection.}
Uniform time sampling tends to return near-duplicate poses, so for Figure~\ref{fig:qual_timeline} we select the $7$ frames that are mutually farthest apart in upper-body joint-angle space (spine through wrists) subject to a minimum temporal separation of $10\%$ of the clip, then display them in temporal order. For the cross-system comparison in Figure~\ref{fig:qual_systems} we deliberately do \emph{not} use this criterion, because choosing timestamps from our own motion would bias the comparison toward moments at which our model happens to move; instead we sample seven evenly spaced timestamps over the middle of the clip and apply the same timestamps to every system. In both figures every tile is cropped with a single shared crop box, computed as the union of the avatar bounding boxes over all frames in the figure, so that the avatar keeps the same scale and image position in every tile and differences between tiles are differences in motion rather than in framing.

\end{document}